\documentclass[prd,twocolumn,showpacs,superscriptaddress,nofootinbib,floatfix,showkeys,10pt]{revtex4-2}

\usepackage{graphicx}
\usepackage{amsmath}
\usepackage{bm}
\usepackage{yhmath}
\usepackage{mathtools}
\usepackage{wasysym}
\usepackage[colorlinks,citecolor=blue,urlcolor=blue,linkcolor=blue]{hyperref}
\usepackage{subfigure}
\usepackage{color}
\usepackage{cases}
\usepackage{subfigure}
\usepackage{times}
\usepackage{dcolumn,booktabs,bm}
\usepackage{slashed}
\usepackage{amsfonts,amssymb,stmaryrd,latexsym}
\usepackage{mathrsfs}
\usepackage{textcomp}
\usepackage{multirow}
\usepackage{cancel}
\usepackage{array}
\usepackage{enumerate,enumitem}
\usepackage{orcidlink}
\usepackage{dashrule}
\usepackage{multirow}
\usepackage{times}
\usepackage{mdframed}

\allowdisplaybreaks[4]
\newcounter{cnt}
\makeatletter
\let\oldhypertarget\hypertarget
\renewcommand{\hypertarget}[2]{%
  \oldhypertarget{#1}{#2}%
    \protected@write\@mainaux{}{%
        \string\expandafter\string\gdef
          \string\csname\string\detokenize{#1}\string\endcsname{#2}%
    }%
  }
\newcommand{\myhyperlink}[1]{%
  \hyperlink{#1}{\csname #1\endcsname}%
  }
\makeatother

\newcommand{\clabel}[2][]{#2}

\def\N2LO{{N$^2$LO}}
\newcommand{\etal}{\textit{et al}.}
\newcommand{\heavy}{\textit{heavy}~}

\begin{document}

\title{Spectrum of the molecular pentaquarks}

\author{Bo Wang\,\orcidlink{0000-0003-0985-2958}}
\affiliation{College of Physics Science \& Technology, Hebei University, Baoding 071002, China}
\affiliation{Hebei Key Laboratory of High-precision Computation and Application of Quantum Field Theory, Baoding, 071002, China}
\affiliation{Hebei Research Center of the Basic Discipline for Computational Physics, Baoding, 071002, China}

\author{Kan Chen\,\orcidlink{0000-0002-1435-6564}}\email{chenk10@nwu.edu.cn}
\affiliation{School of Physics, Northwest University, Xi’an 710127, China}

\author{Lu Meng\,\orcidlink{0000-0001-9791-7138}}\email{lu.meng@rub.de}
\affiliation{Institut f\"ur Theoretische Physik II, Ruhr-Universit\"at Bochum,  D-44780 Bochum, Germany}

\author{Shi-Lin Zhu\,\orcidlink{0000-0002-4055-6906}}\email{zhusl@pku.edu.cn}
\affiliation{School of Physics and Center of High Energy Physics, Peking University, Beijing 100871, China}

\begin{abstract}
We investigate the mass spectrum of the molecular pentaquarks composed of a baryon and a meson. We establish the underlying relations among the near-threshold interactions of the molecular tetraquark and pentaquark systems. We find the existence of the molecule candidates in the $\Sigma_c\bar{D}^{(\ast)}$, $DD^\ast$, and $D\bar{D}^\ast$ systems indicates a substantial presence of the hadronic molecules in the {\it heavy} baryon plus {\it heavy} meson systems ({\it heavy} refers to the hadrons with the $c$ and/or $s$ quarks). We make an exhaustive prediction of the possible bound/virtual molecular states in the systems: $\Sigma_{c}^{(\ast)}\bar{D}^{(\ast)}$, $\Sigma_{c}^{(\ast)}D^{(\ast)}$, $\Xi_{c}^{(\prime,\ast)}\bar{D}^{(\ast)}$, $\Xi_{c}^{(\prime,\ast)}D^{(\ast)}$, $\Sigma_{c}^{(\ast)}K^{\ast}$, $\Sigma_{c}^{(\ast)}\bar{K}^{\ast}$, $\Xi_{c}^{(\prime,\ast)}K^{\ast}$, $\Xi_{c}^{(\prime,\ast)}\bar{K}^{\ast}$, $\Xi_{cc}^{(\ast)}\bar{D}^{(\ast)}$, $\Xi_{cc}^{(\ast)}D^{(\ast)}$, $\Xi_{cc}^{(\ast)}K^{\ast}$, $\Xi_{cc}^{(\ast)}\bar{K}^{\ast}$, $\Sigma^{(\ast)}\bar{D}^{(\ast)}$, $\Sigma^{(\ast)}D^{(\ast)}$, $\Xi^{(\ast)}\bar{D}^{(\ast)}$, $\Xi^{(\ast)}D^{(\ast)}$, $\Sigma^{(\ast)} K^\ast$, $\Sigma^{(\ast)} \bar{K}^\ast$, $\Xi^{(\ast)} K^\ast$, $\Xi^{(\ast)} \bar{K}^\ast$.
Hunting for the predicted states in experiments will significantly deepen our understanding of the formation mechanism of the hadronic molecules, and shed light on the manifestation of flavor symmetry in the low-energy residual strong interactions.

\end{abstract}

\maketitle

\section{Introduction}\label{sec:intro}

An abundance of the exotic near-threshold states has been observed in experiments~\cite{Workman:2022ynf}, starting with the ground-breaking discovery of the $X(3872)$ in 2003~\cite{Belle:2003nnu}. The unconventional nature of these states, characterized by their unique tetraquark and pentaquark compositions, defies the conventional meson and baryon configurations. Unraveling these novel structures has been a vibrant research field of hadron physics over the past two decades. Various theoretical frameworks have been proposed to elucidate these experimental findings, offering different perspectives on the nature of these states. These include the compact multiquark models, kinematic effects, and the existence of the molecular states bound by the residual strong interactions~\cite{Chen:2016qju,Guo:2017jvc,Liu:2019zoy,Lebed:2016hpi,Esposito:2016noz,Brambilla:2019esw,Yang:2020atz,Chen:2021ftn,Chen:2022asf,Meng:2022ozq}.

The LHCb Collaboration made significant breakthroughs in the study of the hidden-charm pentaquarks, starting with the observation of the $P_{\psi}^N(4380)$ and $P_{\psi}^N(4450)$ in the $J/\psi p$ invariant mass spectrum of the $\Lambda_b^0\to J/\psi p K^-$ process in 2015~\cite{LHCb:2015yax}. Four years later, with the updated data, the LHCb identified a new state, $P_{\psi}^N(4312)$, and revealed that the previously reported $P_{\psi}^N(4450)$ actually consists of two distinct states, $P_{\psi}^N(4440)$ and $P_{\psi}^N(4457)$~\cite{LHCb:2019kea}. The $J^P$ quantum numbers of $P_{\psi}^N(4312)$, $P_{\psi}^N(4440)$, and $P_{\psi}^N(4457)$ remain undetermined.
These states are often interpreted as the bound states of $\Sigma_c\bar{D}$ and $\Sigma_c\bar{D}^\ast$, respectively, given that their masses lie below the corresponding thresholds by a few to tens of MeVs~\cite{Meng:2019ilv,Chen:2019bip,Liu:2019tjn,Chen:2019asm,Xiao:2019aya,He:2019ify,Xiao:2019mvs,Guo:2019fdo,Guo:2019kdc,Weng:2019ynv,Burns:2019iih,Wang:2019ato,Du:2019pij,Wang:2019got,Wang:2019spc,Xu:2020gjl,Yalikun:2021bfm,Xing:2022ijm,Zhang:2023czx,Lin:2023ihj}. The discovery of these $P_{\psi}^N$ states has also stimulated predictions for the hidden-charm pentaquarks with strangeness~\cite{Feijoo:2015kts,Chen:2015sxa,Lu:2016roh,Chen:2016ryt,Xiao:2019gjd,Wang:2019nvm}.

More recently, the LHCb Collaboration reported evidence of a new state, $P_{\psi s}^\Lambda(4459)$, near the $\Xi_c\bar{D}^\ast$ threshold in the $J/\psi\Lambda$ invariant mass spectrum from the $\Xi_b^-\to J/\psi\Lambda K^-$ process~\cite{LHCb:2020jpq}. The measured mass of $P_{\psi s}^\Lambda(4459)$ is in excellent agreement with our predictions for the $\Xi_c\bar{D}^\ast$ molecular states in Ref.~\cite{Wang:2019nvm}. The $\Xi_c\bar{D}^\ast$ system is expected to have two states with spin-parity $\frac{1}{2}^-$ and $\frac{3}{2}^-$, respectively. However, the current data sample in Ref.~\cite{LHCb:2020jpq} is insufficient to distinguish between these two structures. Additionally, the LHCb recently observed a state, $P_{\psi s}^\Lambda(4338)$, with high significance in the $J/\psi\Lambda$ invariant mass distribution of the $B^-\to J/\psi\Lambda \bar{p}$ process~\cite{LHCb:2022ogu}. The mass of $P_{\psi s}^\Lambda(4338)$ is very close to the $\Xi_c\bar{D}$ threshold, and experimental analyses favor a spin-parity assignment of $\frac{1}{2}^-$~\cite{LHCb:2022ogu}. The observation of $P_{\psi s}^\Lambda(4459)$ and $P_{\psi s}^\Lambda(4338)$ has sparked intensive studies on their mass spectra~\cite{Chen:2020uif,Peng:2020hql,Wang:2020eep,Liu:2020hcv,Dong:2021juy,Wang:2021hql,Zhu:2021lhd,Xiao:2021rgp,Giron:2021sla,Hu:2021nvs,Chen:2021cfl,Chen:2021spf,Chen:2022onm,Chen:2022wkh}, strong decays~\cite{Chen:2021tip,Lu:2021irg,Yang:2021pio,Azizi:2021pbh}, experimental lineshapes~\cite{Du:2021bgb,Meng:2022wgl,Burns:2022uha,Nakamura:2022gtu}, productions~\cite{Wu:2021caw,Cheng:2021gca,Paryev:2023icm}, and magnetic moments~\cite{Ozdem:2021ugy,Li:2021ryu,Gao:2021hmv,Wang:2022tib,Ozdem:2022kei}.

The LHCb Collaboration has also made remarkable progress in the search for tetraquark states in recent years. The discoveries of the $T_{cs0,1}(2900)$, $T_{cc}(3875)$, and $T_{c\bar{s}0}^a(2900)$ have provided valuable insights into the existence and properties of these exotic hadrons.
The $T_{cs0,1}(2900)$ tetraquark states were observed in the $D^-K^+$ invariant mass spectrum of the $B^+\to D^+D^-K^+$ decay. These states have strangeness and fully open flavor, and their masses are close to the $\bar{D}^\ast K^\ast$ threshold~\cite{LHCb:2020bls,LHCb:2020pxc}. The $T_{cc}(3875)$ is the first observed double-charm tetraquark, discovered in the $D^0D^0\pi^+$ final state~\cite{LHCb:2021vvq,LHCb:2021auc}. This state is a narrow resonance that lies below the $D^{\ast+}D^0$ threshold by a few hundred keV.
More recently, LHCb observed the isovector tetraquark state $T_{c\bar{s}0}^a(2900)$ in the $D_s^+\pi^+$ and $D_s^+\pi^-$ invariant mass distributions of the $B$ decay processes $B^+\to D^-D_s^+\pi^+$ and $B^0\to\bar{D}^0D_s^+\pi^-$~\cite{LHCb:2022sfr,LHCb:2022lzp}. This state was measured to have a spin-parity of $0^+$. Similar to the $T_{cs0}(2900)$, the $T_{c\bar{s}0}^a(2900)$ is located near the $D^\ast K^\ast$ threshold.
These observations of tetraquark states with different flavors and quantum numbers have opened up new avenues for studying and understanding the nature of exotic hadrons.

The near-threshold characteristics of these exotic states suggest that they are highly plausible molecular candidates of the corresponding di-hadron systems. Moreover, these states exhibit a significant isospin tropism, with a majority of them being observed in the lowest isospin channels. The isospin tropism indicates the important role of the one-boson exchange interactions. We expect these molecule candidates do not exist as independent entities, but rather with underlying connections.  We have made significant advancements based on our prior researches~\cite{Meng:2019nzy,Wang:2020dhf,Chen:2021htr}. We have developed a systematic framework that delineates the near-threshold S-wave interactions of the {\it heavy}-light di-hadron systems~\cite{Wang:2023hpp}. In this framework, we assume that the strange quark can be regarded as an inert source or a {\it heavy} quark due to its limited involvement in the near-threshold residual strong interactions. Therefore, the existence of these states is heavily reliant on the correlations of the light quark ($u$ and $d$) pairs between the distinct {\it heavy}-light hadrons. 
This framework categorizes the exotic states into two distinct groups, each governed by the unique light quark pair correlations: the $qq$ type and the $\bar{q}q$ type ($q=u,d$), respectively. Detailed information regarding these interactions can be found in our recent work~\cite{Wang:2023hpp}. Within this comprehensive framework, we have established connections among several observed states, such as $P_{\psi}^N$, $T_{cc}$, and $T_{cs}$, as well as those among $X(3872)$, $Z_c(3900)$, and $T_{c\bar{s}}$~\cite{Wang:2023hpp}. For instance, the recently discovered $T_{cs0}(2900)$ and $T_{c\bar{s}0}^a(2900)$ can be confidently identified as the charm-strange counterparts of $T_{cc}(3875)$ and $Z_c(3900)$, respectively~\cite{Wang:2023hpp}.

In our recent paper~\cite{Wang:2023hpp}, we have successfully constructed the complete mass spectrum of the molecular tetraquarks, involving the $D^{(\ast)}D^{(\ast)}$, $D^{(\ast)}\bar{D}^{(\ast)}$, $D^{(\ast)}K^{\ast}$, and $\bar{D}^{(\ast)}K^{\ast}$ systems. Building upon this progress, we will now extend our calculations to include the spectrum of the molecular pentaquarks, which will encompass additional systems such as $\Sigma_c^{(\ast)}K^\ast$ and $\Sigma_c^{(\ast)}\bar{K}^\ast$, etc. (see Table~\ref{tab:systems} for a comprehensive list of the systems under consideration).

This paper is structured as follows:
Section~\ref{sec:pots} focuses on deriving the effective potentials of the di-hadron systems through the utilization of quark-level potentials. Additionally, we establish the correspondences between the various systems under the framework of {\it heavy} quark symmetry.
In Section~\ref{sec:spec}, we present the calculated mass spectra of the pentaquarks, along with a selection of the important strong decay channels.
Finally, in Section~\ref{sec:sum}, we provide a concise summary of our results and conclude this work.

\section{Near-threshold effective potentials of the di-hadron systems}\label{sec:pots}

As we have elucidated in the Ref.~\cite{Wang:2023hpp}, the near-threshold interactions of two {\it heavy}-light hadrons are dominantly undertaken by the light $qq$ and $\bar{q}q$ ($q=u,d$) correlations. The non-relativistic effective potentials for the light
$qq$ and $\bar{q}q$ can be formulated in the following forms~\cite{Wang:2023hpp}
\begin{eqnarray}
V_{qq}&=&\left(\bm{\tau}_{1}\cdot\bm{\tau}_{2}+\frac{1}{2}\tau_{0,1}\cdot\tau_{0,2}\right)(c_{s}+c_{a}\bm{\sigma}_{1}\cdot\bm{\sigma}_{2}),\label{eq:Vqq}\\
V_{\bar{q}q}&=&\left(-\bm{\tau}_{1}^\ast\cdot\bm{\tau}_{2}+\frac{1}{2}\tau_{0,1}\cdot\tau_{0,2}\right)(\tilde{c}_{s}+\tilde{c}_{a}\bm{\sigma}_{1}\cdot\bm{\sigma}_{2}),\label{eq:Vqqbar}
\end{eqnarray}
where $\tau_0$ denotes the $2\times2$ identity matrix, $\bm\tau$ and $\bm\sigma$ represent the Pauli matrices in the isospin and spin spaces, respectively. The $c_s$ ($\tilde{c}_s$) and $c_a$ ($\tilde{c}_a$) are the low-energy constants (LECs), and their values (ranges) are determined with the well-established states in experiments~\cite{Wang:2023hpp},
\begin{eqnarray}
&c_{s}=146.4\pm10.8\textrm{ GeV}^{-2},&\nonumber\\
&c_{a}=-7.3\pm10.5\textrm{ GeV}^{-2};&\label{eq:csca}\\
&184.3\textrm{ GeV}^{-2}<\tilde{c}_{a}+\tilde{c}_{s}<187.5\textrm{ GeV}^{-2},&\nonumber\\
&78.1\textrm{ GeV}^{-2}<\tilde{c}_{a}-\tilde{c}_{s}<180.3\textrm{ GeV}^{-2}.&
\end{eqnarray}
One can find more comprehensive quark model investigations in Refs.~\cite{Wang:2019rdo,Meng:2023jqk}.

The quark-level potentials $V_{qq}$ ($V_{\bar{q}q}$) can be translated into the hadron-level potentials by sandwiching them between the initial and final di-hadron states. For example, in the case of a system composed of the hadrons $\mathscr{H}_1$ and $\mathscr{H}_2$ (with only $qq$ couplings), the hadron-level potentials can be expressed as
\begin{eqnarray}\label{eq:Vsys}
V_{\mathscr{H}_1\mathscr{H}_2}^{I,J}&=&\left\langle [\mathscr{H}_1\mathscr{H}_2]_J^I\left|V_{qq}\right|[\mathscr{H}_1\mathscr{H}_2]_J^I \right\rangle,
\end{eqnarray}
where the $[\mathscr{H}_1\mathscr{H}_2]_J^I$ represents the spin-flavor wave function of the $\mathscr{H}_1\mathscr{H}_2$ di-hadron systems with the specified total angular momentum $J$ and total isospin $I$~\cite{Chen:2021cfl,Chen:2021spf,Chen:2022wkh}.

\begin{figure}[!ht]
\begin{centering}
    \scalebox{1.0}{\includegraphics[width=\linewidth]{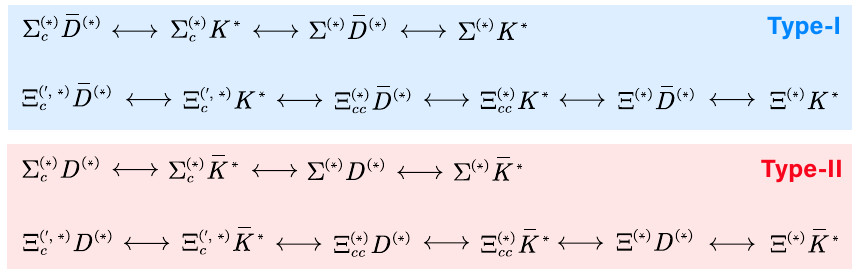}}
    \caption{The correspondences among the systems in Table~\ref{tab:systems} under the {\it heavy} ($c$ and $s$) quark symmetry.\label{fig:correspondences}}
\end{centering}
\end{figure}

The molecular pentaquark systems considered in this study are listed in Table~\ref{tab:systems}. We simultaneously analyze two types of the di-hadron systems, where their interactions are governed by the $qq$ and $\bar{q}q$ couplings respectively. Our previous study has illustrated that the residual strong interactions at the near-threshold energy scale are too weak to excite the strange quark(s) within the hadrons~\cite{Wang:2023hpp}. Therefore, it is reasonable to treat the strange quark as having a similar role to the charm quark. Based on this approximation, we establish the correspondences for the di-hadron systems presented in Table~\ref{tab:systems}, as illustrated in Fig.~\ref{fig:correspondences}.

\begin{table}[!ht]
\centering
\renewcommand{\arraystretch}{1.4}
\caption{The considered pentaquark systems and the corresponding quark contents. The interactions of the Type-I and Type-II systems are assumed to be governed by the $V_{qq}$ and $V_{\bar{q}q}$, respectively.\label{tab:systems}}
\setlength{\tabcolsep}{4.45mm}
{
\begin{tabular}{cccc}
\toprule[0.8pt]
\multicolumn{2}{c}{Type-I: $V_{qq}$} & \multicolumn{2}{c}{Type-II: $V_{\bar{q}q}$}\tabularnewline
\hline 
$\Sigma_{c}^{(\ast)}\bar{D}^{(\ast)}$ & $[cqq][\bar{c}q]$ & $\Sigma_{c}^{(\ast)}D^{(\ast)}$ & $[cqq][c\bar{q}]$\tabularnewline
$\Xi_{c}^{(\prime,\ast)}\bar{D}^{(\ast)}$ & $[csq][\bar{c}q]$ & $\Xi_{c}^{(\prime,\ast)}D^{(\ast)}$ & $[csq][c\bar{q}]$\tabularnewline
$\Sigma_{c}^{(\ast)}K^{\ast}$ & $[cqq][\bar{s}q]$ & $\Sigma_{c}^{(\ast)}\bar{K}^{\ast}$ & $[cqq][s\bar{q}]$\tabularnewline
$\Xi_{c}^{(\prime,\ast)}K^{\ast}$ & $[csq][\bar{s}q]$ & $\Xi_{c}^{(\prime,\ast)}\bar{K}^{\ast}$ & $[csq][s\bar{q}]$\tabularnewline
$\Xi_{cc}^{(\ast)}\bar{D}^{(\ast)}$ & $[ccq][\bar{c}q]$ & $\Xi_{cc}^{(\ast)}D^{(\ast)}$ & $[ccq][c\bar{q}]$\tabularnewline
$\Xi_{cc}^{(\ast)}K^{\ast}$ & $[ccq][\bar{s}q]$ & $\Xi_{cc}^{(\ast)}\bar{K}^{\ast}$ & $[ccq][s\bar{q}]$\tabularnewline
$\Sigma^{(\ast)}\bar{D}^{(\ast)}$ & $[sqq][\bar{c}q]$ & $\Sigma^{(\ast)}D^{(\ast)}$ & $[sqq][c\bar{q}]$\tabularnewline
$\Xi^{(\ast)}\bar{D}^{(\ast)}$ & $[ssq][\bar{c}q]$ & $\Xi^{(\ast)}D^{(\ast)}$ & $[ssq][c\bar{q}]$\tabularnewline
$\Sigma^{(\ast)}K^\ast$ & $[sqq][\bar{s}q]$ & $\Sigma^{(\ast)}\bar{K}^\ast$ & $[sqq][s\bar{q}]$\tabularnewline
$\Xi^{(\ast)}K^\ast$ & $[ssq][\bar{s}q]$ & $\Xi^{(\ast)}\bar{K}^\ast$ & $[ssq][s\bar{q}]$\tabularnewline
\bottomrule[0.8pt]
\end{tabular}
}
\end{table}

The effective potentials for the $\Sigma_c^{(\ast)}\bar{D}^{(\ast)}$ and $\Xi_c^{(\prime,\ast)}\bar{D}^{(\ast)}$ systems are presented in Table~\ref{tab:potentials}. To obtain the effective potentials for the corresponding $\Sigma_c^{(\ast)}D^{(\ast)}$ and $\Xi_c^{(\prime,\ast)}D^{(\ast)}$ systems in Type-II, the following replacements should be made:
\begin{eqnarray}
c_s\to \tilde{c}_s,\qquad c_a\to \tilde{c}_a.
\end{eqnarray}
For the other systems in Type-I(II), the effective potentials can be directly obtained using the correspondences depicted in Fig.~\ref{fig:correspondences}. For example, for the Type-I systems, $V_{\Sigma_c \bar{D}^\ast}^{I,J}=V_{\Sigma_c K^\ast}^{I,J}=V_{\Sigma \bar{D}^\ast}^{I,J}=V_{\Sigma K^\ast}^{I,J}$, and $V_{\Xi_c^\prime\bar{D}^\ast}^{I,J}=V_{\Xi_{c}^\prime K^\ast}^{I,J}=V_{\Xi_{cc}\bar{D}^\ast}^{I,J}=V_{\Xi_{cc}K^\ast}^{I,J}=V_{\Xi\bar{D}^\ast}^{I,J}=V_{\Xi K^\ast}^{I,J}$. Similar relations hold for the Type-II systems. The matrix elements of the isospin-isospin and spin-spin coupling operators for the di-hadrons with a definite isospin $I$ and angular momentum $J$ are listed in Table~\ref{tab:IdotI}, which allows for the projection of the effective potentials onto a specific $I(J^P)$ state.

\begin{table}[!ht]
\centering
\renewcommand{\arraystretch}{1.5}
\caption{The effective potentials of the Type-I systems, where $\mathbf{I}_1$ ($\mathbf{S}_{1}$) and $\mathbf{I}_{2}$ ($\mathbf{S}_{2}$) denote the isospin (spin) of the {\it heavy} baryons and {\it heavy} mesons, respectively.\label{tab:potentials}}
\setlength{\tabcolsep}{0.8mm}
{
\begin{tabular}{ccc}
\toprule[0.8pt]
Systems & Effective potentials & $|\mathbf{I}_{1}\otimes\mathbf{I}_{2}\rangle,|\mathbf{S}_{1}\otimes\mathbf{S}_{2}\rangle$\tabularnewline
\hline 
$\Sigma_{c}\bar{D}$ & $\left(1+4\mathbf{I}_{1}\cdot\mathbf{I}_{2}\right)c_{s}$ & $|1\otimes\frac{1}{2}\rangle,|\frac{1}{2}\otimes0\rangle$\tabularnewline
$\Sigma_{c}\bar{D}^{\ast}$ & $\left(1+4\mathbf{I}_{1}\cdot\mathbf{I}_{2}\right)\left(c_{s}+\frac{4}{3}c_{a}\mathbf{S}_{1}\cdot\mathbf{S}_{2}\right)$ & $|1\otimes\frac{1}{2}\rangle,|\frac{1}{2}\otimes1\rangle$\tabularnewline
$\Sigma_{c}^{\ast}\bar{D}$ & $\left(1+4\mathbf{I}_{1}\cdot\mathbf{I}_{2}\right)c_{s}$ & $|1\otimes\frac{1}{2}\rangle,|\frac{3}{2}\otimes0\rangle$\tabularnewline
$\Sigma_{c}^{\ast}\bar{D}^{\ast}$ & $\left(1+4\mathbf{I}_{1}\cdot\mathbf{I}_{2}\right)\left(c_{s}+\frac{2}{3}c_{a}\mathbf{S}_{1}\cdot\mathbf{S}_{2}\right)$ & $|1\otimes\frac{1}{2}\rangle,|\frac{3}{2}\otimes1\rangle$\tabularnewline
$\Xi_{c}\bar{D}$ & $\left(\frac{1}{2}+4\mathbf{I}_{1}\cdot\mathbf{I}_{2}\right)c_{s}$ & $|\frac{1}{2}\otimes\frac{1}{2}\rangle,|\frac{1}{2}\otimes0\rangle$\tabularnewline
$\Xi_{c}\bar{D}^{\ast}$ & $\left(\frac{1}{2}+4\mathbf{I}_{1}\cdot\mathbf{I}_{2}\right)\left(c_{s}+2c_{a}\mathbf{S}_{1}\cdot\mathbf{S}_{2}\right)$ & $|\frac{1}{2}\otimes\frac{1}{2}\rangle,|\frac{1}{2}\otimes1\rangle$\tabularnewline
$\Xi_{c}^{\prime}\bar{D}$ & $\left(\frac{1}{2}+4\mathbf{I}_{1}\cdot\mathbf{I}_{2}\right)c_{s}$ & $|\frac{1}{2}\otimes\frac{1}{2}\rangle,|\frac{1}{2}\otimes0\rangle$\tabularnewline
$\Xi_{c}^{\prime}\bar{D}^{\ast}$ & $\left(\frac{1}{2}+4\mathbf{I}_{1}\cdot\mathbf{I}_{2}\right)\left(c_{s}-\frac{2}{3}c_{a}\mathbf{S}_{1}\cdot\mathbf{S}_{2}\right)$ & $|\frac{1}{2}\otimes\frac{1}{2}\rangle,|\frac{1}{2}\otimes1\rangle$\tabularnewline
$\Xi_{c}^{\ast}\bar{D}$ & $\left(\frac{1}{2}+4\mathbf{I}_{1}\cdot\mathbf{I}_{2}\right)c_{s}$ & $|\frac{1}{2}\otimes\frac{1}{2}\rangle,|\frac{3}{2}\otimes0\rangle$\tabularnewline
$\Xi_{c}^{\ast}\bar{D}^{\ast}$ & $\left(\frac{1}{2}+4\mathbf{I}_{1}\cdot\mathbf{I}_{2}\right)\left(c_{s}+\frac{2}{3}c_{a}\mathbf{S}_{1}\cdot\mathbf{S}_{2}\right)$ & $|\frac{1}{2}\otimes\frac{1}{2}\rangle,|\frac{3}{2}\otimes1\rangle$\tabularnewline
\bottomrule[0.8pt]
\end{tabular}
}
\end{table}

\begin{table}[!ht]
\centering
\renewcommand{\arraystretch}{1.4}
\caption{Matrix elements of the isospin-isospin and spin-spin coupling operators in Table~\ref{tab:potentials}.\label{tab:IdotI}}
\setlength{\tabcolsep}{4.5mm}
{
\begin{tabular}{cccccc}
\toprule[0.8pt]
$|\mathbf{I}_{1}\otimes\mathbf{I}_{2}\rangle$ & \multicolumn{2}{c}{$|\frac{1}{2}\otimes\frac{1}{2}\rangle$} & \multicolumn{2}{c}{$|\frac{1}{2}\otimes1\rangle$} & \tabularnewline
\hline 
$I$ & $0$ & $1$ & $\frac{1}{2}$ & $\frac{3}{2}$ & \tabularnewline
$\langle\mathbf{I}_{1}\cdot\mathbf{I}_{2}\rangle$ & $-\frac{3}{4}$ & $\frac{1}{4}$ & $-1$ & $\frac{1}{2}$ & \tabularnewline
\hline 
$|\mathbf{S}_{1}\otimes\mathbf{S}_{2}\rangle$ & \multicolumn{2}{c}{$|\frac{1}{2}\otimes1\rangle$} & \multicolumn{3}{c}{$|\frac{3}{2}\otimes1\rangle$}\tabularnewline
\hline 
$J$ & $\frac{1}{2}$ & $\frac{3}{2}$ & $\frac{1}{2}$ & $\frac{3}{2}$ & $\frac{5}{2}$\tabularnewline
$\langle\mathbf{S}_{1}\cdot\mathbf{S}_{2}\rangle$ & $-1$ & $\frac{1}{2}$ & $-\frac{5}{2}$ & $-1$ & $\frac{3}{2}$\tabularnewline
\bottomrule[0.8pt]
\end{tabular}
}
\end{table}

In this study, we exclude the systems that involve an isospin singlet $\Lambda$ or $\Lambda_c$ baryon. Previous researches have shown that the interactions in the systems such as $\Lambda_c\bar{D}^{(\ast)}$ are feeble~\cite{Wang:2011rga,Yang:2011wz,Wang:2019nvm} due to the absence of the isospin couplings. In our calculations, we have $V_{\Lambda_c\bar{D}^{(\ast)}}=c_s>0$, which means that no bound states can be formed in these systems.

To search for the possible bound or virtual states, we solve the Lippmann-Schwinger equations (LSEs) given by
\begin{eqnarray}\label{eq:intLSE}
t=v+vGt,
\end{eqnarray}
where $v$ denotes the $V_{\mathscr{H}_1\mathscr{H}_2}^{I,J}$ derived above, and $G$ is the non-relativistic two-body propagator. The explicit form of $G$ is given by
\begin{eqnarray}\label{eq:Ge}
  G(E+i\epsilon) &=& \int_{0}^{\Lambda}\frac{k^{2}dk}{(2\pi)^{3}}\frac{2\mu}{p^{2}-k^{2}+i\epsilon}\nonumber\\
    &=& \frac{2\mu}{(2\pi)^{3}}\left[p\tanh^{-1}\left(\frac{p}{\Lambda}\right)-\Lambda-\frac{i\pi}{2}p\right],
\end{eqnarray}
where a sharp cutoff $\Lambda$ is introduced to regularize the loop integral. We will use $\Lambda=0.4$ GeV, as in Ref.~\cite{Wang:2023hpp}. $\mu$ represents the reduced mass of the corresponding two-body system, and $p=[2\mu(E-m_{\text{th}})]^{1/2}$, where $E$ is the center-of-mass energy and $m_{\text{th}}$ is the threshold.

The integral equation~\eqref{eq:intLSE} can be reduced to an algebraic equation for the energy-independent effective potentials,
\begin{eqnarray}\label{eq:invt}
  t^{-1} &=& v^{-1}-G.
\end{eqnarray}
\clabel[im]{}The physical T-matrix, as a function of the center-of-mass energy $E$, is given by Eq.~\eqref{eq:intLSE} or~\eqref{eq:invt}. However, in order to investigate the positions or distributions of poles in the T-matrix, one must consider the analytical structure of the scattering T-matrix on different Riemann sheets. For the case of single-channel scattering, the branch cut of the propagator in Eq.~\eqref{eq:Ge} gives rise to two Riemann sheets, known as the physical sheet and the unphysical sheet, respectively, when $E > m_{\mathrm{th}}$. When transitioning from the physical sheet to the unphysical sheet, one needs to cross the branch cut, which leads to a so-called discontinuity primarily caused by the different signs of the imaginary part of $G$, i.e., $\Im G(E+i\epsilon)=-\Im G(E-i\epsilon)$. Therefore, an analytic continuation must be performed to ensure the continuity of the propagator $G$ when crossing the branch cut. If we use I and II to denote the physical and unphysical sheets, respectively, the corresponding $G$ must satisfy the following conditions:
\begin{eqnarray}
G_{\mathrm{II}}(E+i\epsilon)=G_{\mathrm{I}}(E-i\epsilon).
\end{eqnarray}
Then we have
\begin{eqnarray}
G_{\mathrm{II}}(E+i\epsilon)=G_{\mathrm{I}}(E-i\epsilon)=G_{\mathrm{I}}(E+i\epsilon)-2i\Im G_{\mathrm{I}}(E+i\epsilon).\nonumber\\
\end{eqnarray}
From the Eq.~\eqref{eq:Ge} one obtains that $\Im G_{\mathrm{I}}(E+i\epsilon)=-\mu p/(8\pi)^2$.

The bound and virtual states exist as the zeros of the inverse T-matrix $t^{-1}$ in the corresponding physical and unphysical Riemann sheets, respectively. Then the physical and unphysical sheets are associated with the following replacements for the $G$ function:
\begin{eqnarray}\label{eq:phyunphy}
\text{Physical}&:& G(E+i\epsilon),\nonumber\\
\text{Unphysical}&:& G(E+i\epsilon)+i\frac{\mu}{4\pi^2} p,
\end{eqnarray}
where the $G(E+i\epsilon)$ is given by Eq.~\eqref{eq:Ge}.

\section{Spectrum of the molecular pentaquarks}\label{sec:spec}

\subsection{$\Sigma_c^{(\ast)}\bar{D}^{(\ast)}$, $\Xi_c^{(\prime,\ast)}\bar{D}^{(\ast)}$ and $\Sigma_c^{(\ast)}D^{(\ast)}$, $\Xi_c^{(\prime,\ast)}D^{(\ast)}$ systems}\label{sec:specSD}

\begin{figure*}[!ht]
\begin{centering}
    \scalebox{1.0}{\includegraphics[width=\linewidth]{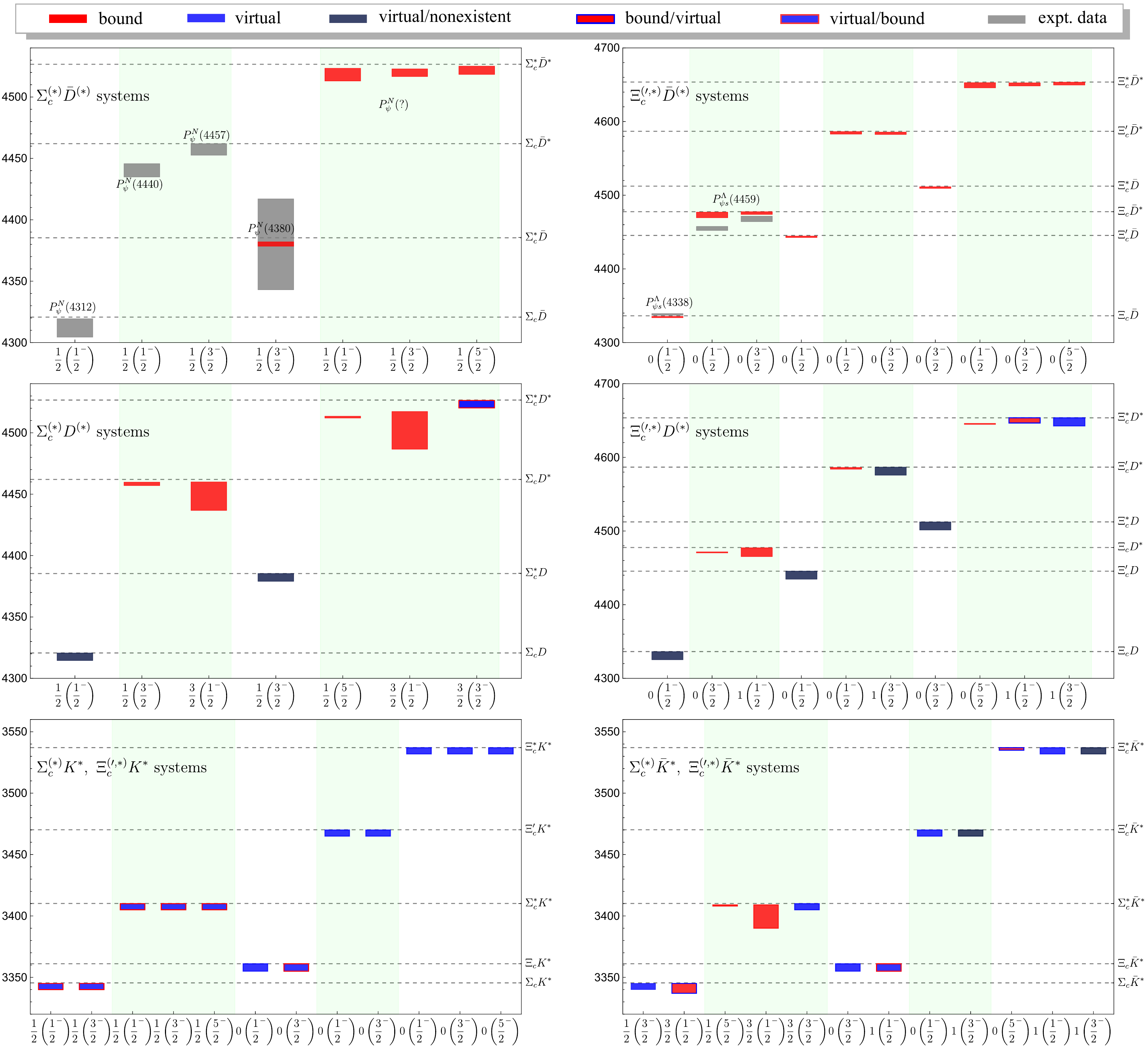}}
    \caption{Spectrum of the molecular pentaquarks with a charmed baryon. The $x$- and $y$-axes represent the $I(J^P)$ quantum numbers and the masses (in units of MeV), respectively. The corresponding di-hadron systems are marked in each subfigure. We use the red, blue and black rectangles to denote the bound states, virtual states and the possibly nonexistent states, respectively, while the red (blue) rectangles with blue (red) edges means that the states can also be the near-threshold virtual (bound) states. The dark gray rectangles represent the experimental data. We stress that only the vertical widthes of red rectangles qualitatively reflect the corresponding mass ranges. The corresponding thresholds are denoted with the horizontal dashed lines. The experimental data for $P_{\psi}^N$ and $P_{\psi s}^\Lambda$ states are taken from Refs.~\cite{LHCb:2019kea,LHCb:2020jpq,LHCb:2022ogu}, where the results with two-peak hypothesis for $P_{\psi s}^\Lambda(4459)$~\cite{LHCb:2020jpq} are quoted.\label{fig:spectrum1}}
\end{centering}
\end{figure*}

The mass spectra and strong decay patterns of the molecular pentaquarks in the $\Sigma_c^{(\ast)}\bar{D}^{(\ast)}$, $\Xi_c^{(\prime,\ast)}\bar{D}^{(\ast)}$, $\Sigma_c^{(\ast)}D^{(\ast)}$, and $\Xi_c^{(\prime,\ast)}D^{(\ast)}$ systems are presented in Table~\ref{tab:specSD}. These results are also summarized in the first two rows of Fig.~\ref{fig:spectrum1}, which illustrate the mass spectra and relative positions of the relevant thresholds. In the subsequent analysis, we will examine the outcomes of these systems in a systematic manner.

The three $P_{\psi}^N$ states within the $\Sigma_c\bar{D}^{(\ast)}$ systems are considered as the inputs in our study~\cite{Wang:2023hpp}. We predict the existence of four bound states in the isospin-$\frac{1}{2}$ channels of the $\Sigma_c^\ast\bar{D}^{(\ast)}$ systems, which is consistent with the findings of several previous investigations~\cite{Chen:2019bip,Liu:2019tjn,Xiao:2019aya,Wang:2019ato,Du:2019pij,He:2019rva}. Notably, the bound state with a mass around $4380.3$ MeV in the $\Sigma_c^\ast\bar{D}$ system shall correspond to the previously reported $P_{\psi}^N(4380)$~\cite{LHCb:2015yax}. However, no molecular states are found in the isospin-$\frac{3}{2}$ channels. The discovery of the $P_{\psi}^N$ states has sparked significant interest in the investigation of their mass spectra, strong decays, production mechanisms, and experimental line-shapes, leading to a plethora of studies in this field (see the recent review~\cite{Meng:2022ozq}).

In total, there are ten bound states identified in the isoscalar $\Xi_c^{(\prime,\ast)}\bar{D}^{(\ast)}$ systems, which is consistent with our previous predictions~\cite{Wang:2019nvm}. In Ref.~\cite{Wang:2019nvm}, the mass of the $\Xi_c\bar{D}$ bound state was estimated to be approximately $4320$ MeV, about $15$ MeV lower than the $\Xi_c\bar{D}$ threshold. However, the recent observation of the $P_{\psi s}^\Lambda(4338)$ state indicates that it is very close to the $\Xi_c\bar{D}$ threshold~\cite{LHCb:2022ogu}. This discrepancy can be naturally resolved within the framework of {\it heavy} quark symmetry. By considering this symmetry, the $\Xi_c\bar{D}$ interaction is governed by a single pair $qq$ coupling, similar to the $DD^\ast$ system, where the $T_{cc}$ state exists. From Table~\ref{tab:potentials} and Ref.~\cite{Wang:2023hpp}, the potentials for the isoscalar $\Xi_c\bar{D}$ and $DD^\ast$ respectively read\footnote{The contribution from crossed diagrams in $DD^\ast\to DD^\ast$ scattering will give rise to an effective spin transition between $D$ and $D^\ast$, resulting in an additional $c_a$ term in the $DD^\ast$ potentials.}
\begin{eqnarray}
V_{\Xi_c\bar{D}}^{0,1/2}=-\frac{5}{2}c_s,\qquad V_{DD^\ast}^{0,1}=-\frac{5}{2}c_s+\frac{5}{2}c_a.
\end{eqnarray}
Considering $|c_s|\gg|c_a|$ [e.g., see the Eq.~\eqref{eq:csca}], it is evident that $V_{\Xi_c\bar{D}}^{0,1/2}\approx V_{DD^\ast}^{0,1}$. Thus, a straightforward connection can be established between the $T_{cc}(3875)$ and $P_{\psi s}^\Lambda(4338)$ states, and their near-threshold characteristics can be simultaneously explained within a unified framework. Additionally, it is worth noting that the other bound states in the $\Xi_c^{(\prime,\ast)}\bar{D}^{(\ast)}$ systems are characterized by weak binding energies, typically a few MeVs. The $0(\frac{1}{2}^-)$ or (and) $0(\frac{3}{2}^-)$ state(s) in the $\Xi_c\bar{D}^\ast$ system is expected to correspond to the recently observed signal of the $P_{\psi s}^\Lambda(4459)$~\cite{LHCb:2020jpq}.

A series of studies have been conducted on the possible existence of double-charm pentaquarks, following the discovery of the $P_{\psi}^N$ states~\cite{Shimizu:2017xrg,Wang:2018lhz,Zhou:2018bkn,Park:2018oib,Yang:2020twg,Chen:2021htr,Dong:2021bvy,Chen:2021kad,Shen:2022zvd,Wang:2023aob}. For instance, Chen \etal~investigated the spectrum of $\Sigma_c^{(\ast)}D^{(\ast)}$ molecules using the chiral effective field theory and found a similar pattern to that of the $P_{\psi}^N$ states, with seven bound states in the isospin-$\frac{1}{2}$ channels~\cite{Chen:2021htr}. Similar results were also obtained in Ref.~\cite{Dong:2021bvy}. By employing the one-boson exchange (OBE) model, several isospin-$\frac{1}{2}$ states in the coupled systems of $\Sigma_c^{(\ast)}D^{(\ast)}$ were obtained in Refs.~\cite{Shimizu:2017xrg,Chen:2021kad,Shen:2022zvd}. Additionally, the chiral quark model was utilized to predict several narrow resonances in $\Sigma_c^{(\ast)}D^{(\ast)}$ systems~\cite{Yang:2020twg,Liu:2023clr}.

Our calculations indicate that the spectrum of $\Sigma_c^{(\ast)}D^{(\ast)}$ differs significantly from that of $\Sigma_c^{(\ast)}\bar{D}^{(\ast)}$. Notably, not all isospin-$\frac{1}{2}$ channels can form molecular states (see Table~\ref{tab:specSD} and Fig.~\ref{fig:spectrum1}). The most promising candidates are the $\frac{1}{2}(\frac{3}{2}^-)$ and $\frac{3}{2}(\frac{1}{2}^-)$ states in $\Sigma_c D^\ast$, as well as the $\frac{1}{2}(\frac{5}{2}^-)$, $\frac{3}{2}(\frac{1}{2}^-)$, and $\frac{3}{2}(\frac{3}{2}^-)$ states in $\Sigma_c^\ast D^\ast$. The two isospin-$\frac{1}{2}$ states in $\Sigma_c^{(\ast)}D$ systems may exist as virtual states or disappear when $|\tilde{c}_s|$ is small.

We also examined the spectrum of double-charm pentaquarks composed of $\Xi_c^{(\prime,\ast)}$ and $D^{(\ast)}$, which can be regarded as the double-charm counterparts of the $P_{\psi s}^\Lambda$ states. Our analysis reveals that the $0(\frac{3}{2}^-)$ and $1(\frac{1}{2}^-)$ states in $\Xi_c D^\ast$, the $0(\frac{1}{2}^-)$ state in $\Xi_c^\prime D^\ast$, as well as the $0(\frac{5}{2}^-)$, $1(\frac{1}{2}^-)$, and $1(\frac{3}{2}^-)$ states in $\Xi_c^\ast D^\ast$ are promising candidates of the molecular states. However, the isoscalar states in $\Xi_c^{(\prime,\ast)}D$ systems may exist as virtual states or not exist at all. Recently, Wang \etal~also examined the interactions of $\Xi_c^{(\prime,\ast)}D^{(\ast)}$ systems using the OBE model and identified ten hadronic molecules in the isoscalar channels~\cite{Wang:2023ael}. The spectrum obtained in Ref.~\cite{Wang:2023ael} follows a similar pattern to that of the $P_{\psi s}^\Lambda$ spectrum.

It is worth noting that the double-charm molecular pentaquarks in $\Sigma_c^{(\ast)}D^{(\ast)}$ and $\Xi_c^{(\prime,\ast)}D^{(\ast)}$ systems can decay into the double-charm baryons [$\Xi_{cc}^{(\ast)},\Omega_{cc}^{(\ast)}$] in combination with light hadrons ($\pi,\rho,\omega,\dots$). Therefore, these hadronic molecules have the potential to serve as the sources of the production of the double-charm baryons.

\begin{table*}[!ht]
\centering
\renewcommand{\arraystretch}{1.4}
\caption{The mass spectra and selected decay channels of the molecular pentaquarks in $\Sigma_c^{(\ast)}\bar{D}^{(\ast)}$, $\Xi_c^{(\prime,\ast)}\bar{D}^{(\ast)}$ and $\Sigma_c^{(\ast)}D^{(\ast)}$, $\Xi_c^{(\prime,\ast)}D^{(\ast)}$ systems. The isospin average masses of the corresponding systems are given in the brackets in the second column, e.g., in the form $\Sigma_{c}\bar{D}~[4320.7]$. The subscripts $B/V$ denote the bound/virtual state solutions. The superscript $\sharp$ means that this state might be nonexistent in the range of parameters, while the $\dagger$ represents that this state can also be the near-threshold bound (virtual) state if it is labeled as the virtual (bound) state in this table. The (pole) masses are given in units of MeV.\label{tab:specSD}}
\setlength{\tabcolsep}{1.9mm}
{
\begin{tabular}{ccccccc}
\toprule[0.8pt]
$I(J^{P})$ & Systems [Thr.] & $E_{B}/E_{V}$ & Decay channels & Systems & $E_{B}/E_{V}$ & Decay channels\tabularnewline
\hline 
$\frac{1}{2}\left(\frac{1}{2}^{-}\right)$ & \multirow{2}{*}{$\Sigma_{c}\bar{D}~[4320.7]$} & $\left[4311.9\pm0.7_{-0.6}^{+6.8}\right]_{B}$ & $J/\psi N,\eta_{c}N,\Lambda_{c}\bar{D}^{(\ast)}$ & \multirow{2}{*}{$\Sigma_{c}D$} & $\left[\sim4308.4\right]_{V}^{\sharp}$ & $\Lambda_{c}D^{(\ast)},\Xi_{cc}\pi$\tabularnewline
$\frac{3}{2}\left(\frac{1}{2}^{-}\right)$ &  & $-$ & $-$ &  & $-$ & $-$\tabularnewline
\hline 
$\frac{1}{2}\left(\frac{1}{2}^{-}\right)$ & \multirow{4}{*}{$\Sigma_{c}\bar{D}^{\ast}~[4462.0]$} & $\left[4440.3\pm1.3_{-4.7}^{+4.1}\right]_{B}$ & $J/\psi N,\eta_{c}N,\Lambda_{c}\bar{D}^{(\ast)}$ & \multirow{4}{*}{$\Sigma_{c}D^{\ast}$} & $-$ & $-$\tabularnewline
$\frac{1}{2}\left(\frac{3}{2}^{-}\right)$ &  & $\left[4457.3\pm0.6_{-1.7}^{+4.1}\right]_{B}$ & $J/\psi N,\Lambda_{c}\bar{D}^{\ast}$ &  & $\left[4457.0,4459.8\right]_{B}$ & $\Lambda_{c}D^{\ast},\Xi_{cc}^{\ast}\pi,\Xi_{cc}\rho(\omega)$\tabularnewline
$\frac{3}{2}\left(\frac{1}{2}^{-}\right)$ &  & $-$ & $-$ &  & $\left[4436.8,4460.0\right]_{B}$ & $\Sigma_{c}D,\Xi_{cc}\pi,\Xi_{cc}\rho$\tabularnewline
$\frac{3}{2}\left(\frac{3}{2}^{-}\right)$ &  & $-$ & $-$ &  & $-$ & $-$\tabularnewline
\hline 
$\frac{1}{2}\left(\frac{3}{2}^{-}\right)$ & \multirow{2}{*}{$\Sigma_{c}^{\ast}\bar{D}~[4385.6]$} & $\left[4380.3_{-1.8}^{+1.7}\right]_{B}$ & $J/\psi N,\Lambda_{c}\bar{D}^{\ast}$ & \multirow{2}{*}{$\Sigma_{c}^{\ast}D$} & $\left[\sim4373.7\right]_{V}^{\sharp}$ & $\Lambda_{c}D^{\ast},\Xi_{cc}^{\ast}\pi,\Xi_{cc}\rho(\omega)$\tabularnewline
$\frac{3}{2}\left(\frac{3}{2}^{-}\right)$ &  & $-$ & $-$ &  & $-$ & $-$\tabularnewline
\hline 
$\frac{1}{2}\left(\frac{1}{2}^{-}\right)$ & \multirow{6}{*}{$\Sigma_{c}^{\ast}\bar{D}^{\ast}~[4526.7]$} & $\left[4518.7_{-5.6}^{+4.7}\right]_{B}$ & $J/\psi N,\eta_{c}N,\Lambda_{c}\bar{D}^{(\ast)}$ & \multirow{6}{*}{$\Sigma_{c}^{\ast}D^{\ast}$} & $-$ & $-$\tabularnewline
$\frac{1}{2}\left(\frac{3}{2}^{-}\right)$ &  & $\left[4520.0_{-3.3}^{+2.9}\right]_{B}$ & $J/\psi N,\Lambda_{c}\bar{D}^{\ast}$ &  & $-$ & $-$\tabularnewline
$\frac{1}{2}\left(\frac{5}{2}^{-}\right)$ &  & $\left[4522.1_{-3.7}^{+3.0}\right]_{B}$ & $J/\psi N,\eta_{c}N,\Lambda_{c}\bar{D}^{(\ast)}$ &  & $\left[4513.0,4513.5\right]_{B}$ & $\Lambda_{c}D^{(\ast)},\Xi_{cc}^{(\ast)}\pi,\Xi_{cc}^{(\ast)}\rho(\omega)$\tabularnewline
$\frac{3}{2}\left(\frac{1}{2}^{-}\right)$ &  & $-$ & $-$ &  & $\left[4486.6,4517.3\right]_{B}$ & $\Sigma_{c}D,\Xi_{cc}\pi,\Xi_{cc}\rho$\tabularnewline
$\frac{3}{2}\left(\frac{3}{2}^{-}\right)$ &  & $-$ & $-$ &  & $\left[4483.8,4526.3\right]_{V}^{\dagger}$ & $\Sigma_{c}D^{\ast},\Xi_{cc}^{\ast}\pi,\Xi_{cc}\rho$\tabularnewline
$\frac{3}{2}\left(\frac{5}{2}^{-}\right)$ &  & $-$ & $-$ &  & $-$ & $-$\tabularnewline
\hline 
$0\left(\frac{1}{2}^{-}\right)$ & \multirow{2}{*}{$\Xi_{c}\bar{D}~[4336.3]$} & $\left[4334.8_{-1.1}^{+0.8}\right]_{B}$ & $J/\psi\Lambda,\eta_{c}\Lambda,\bar{D}_{s}\Lambda_{c}$ & \multirow{2}{*}{$\Xi_{c}D$} & $\left[\sim4313.5\right]_{V}^{\sharp}$ & $\Xi_{cc}\bar{K},\Omega_{cc}\pi^{+}\pi^{-}\pi^{0}$\tabularnewline
$1\left(\frac{1}{2}^{-}\right)$ &  & $-$ & $-$ &  & $-$ & $-$\tabularnewline
\hline 
$0\left(\frac{1}{2}^{-}\right)$ & \multirow{4}{*}{$\Xi_{c}\bar{D}^{\ast}~[4477.6]$} & $\left[4473.9_{-4.5}^{+3.1}\right]_{B}$ & $J/\psi\Lambda,\eta_{c}\Lambda,\bar{D}_{s}^{(\ast)}\Lambda_{c}$ & \multirow{4}{*}{$\Xi_{c}D^{\ast}$} & $-$ & $-$\tabularnewline
$0\left(\frac{3}{2}^{-}\right)$ &  & $\left[4476.3_{-2.3}^{+1.3}\right]_{B}$ & $J/\psi\Lambda,\bar{D}_{s}^{\ast}\Lambda_{c}$ &  & $\left[4470.3,4470.7\right]_{B}$ & $\Xi_{cc}^{\ast}\bar{K},\Omega_{cc}\pi^{+}\pi^{-}\pi^{0}$\tabularnewline
$1\left(\frac{1}{2}^{-}\right)$ &  & $-$ & $-$ &  & $\left[4465.3,4477.2\right]_{B}$ & $\Xi_{cc}\bar{K},\Omega_{cc}\pi$\tabularnewline
$1\left(\frac{3}{2}^{-}\right)$ &  & $-$ & $-$ &  & $-$ & $-$\tabularnewline
\hline 
$0\left(\frac{1}{2}^{-}\right)$ & \multirow{2}{*}{$\Xi_{c}^{\prime}\bar{D}~[4445.6]$} & $\left[4443.7_{-1.1}^{+0.9}\right]_{B}$ & $J/\psi\Lambda,\eta_{c}\Lambda,\bar{D}_{s}^{(\ast)}\Lambda_{c}$ & \multirow{2}{*}{$\Xi_{c}^{\prime}D$} & $\left[\sim4424.4\right]_{V}^{\sharp}$ & $\Xi_{cc}\bar{K},\Omega_{cc}\pi^{+}\pi^{-}\pi^{0}$\tabularnewline
$1\left(\frac{1}{2}^{-}\right)$ &  & $-$ & $-$ &  & $-$ & $-$\tabularnewline
\hline 
$0\left(\frac{1}{2}^{-}\right)$ & \multirow{4}{*}{$\Xi_{c}^{\prime}\bar{D}^{\ast}~[4586.8]$} & $\left[4585.0_{-2.0}^{+1.4}\right]_{B}$ & $J/\psi\Lambda,\eta_{c}\Lambda,\bar{D}_{s}^{(\ast)}\Lambda_{c}$ & \multirow{4}{*}{$\Xi_{c}^{\prime}D^{\ast}$} & $\left[4584.8,4586.3\right]_{B}$ & $\Xi_{c}D,\Xi_{cc}\bar{K},\Omega_{cc}\omega$\tabularnewline
$0\left(\frac{3}{2}^{-}\right)$ &  & $\left[4584.2_{-1.7}^{+1.4}\right]_{B}$ & $J/\psi\Lambda,\bar{D}_{s}^{\ast}\Lambda_{c}$ &  & $-$ & $-$\tabularnewline
$1\left(\frac{1}{2}^{-}\right)$ &  & $-$ & $-$ &  & $-$ & $-$\tabularnewline
$1\left(\frac{3}{2}^{-}\right)$ &  & $-$ & $-$ &  & $\left[\sim4526.5\right]_{V}^{\sharp}$ & $\Xi_{cc}^{\ast}\bar{K},\Omega_{cc}^{\ast}\pi$\tabularnewline
\hline 
$0\left(\frac{3}{2}^{-}\right)$ & \multirow{2}{*}{$\Xi_{c}^{\ast}\bar{D}~[4512.4]$} & $\left[4510.5_{-1.2}^{+0.9}\right]_{B}$ & $J/\psi\Lambda,\bar{D}_{s}^{\ast}\Lambda_{c}$ & \multirow{2}{*}{$\Xi_{c}^{\ast}D$} & $\left[\sim4492.2\right]_{V}^{\sharp}$ & $\Xi_{cc}^{\ast}\bar{K},\Omega_{cc}\pi^{+}\pi^{-}\pi^{0}$\tabularnewline
$1\left(\frac{3}{2}^{-}\right)$ &  & $-$ & $-$ &  & $-$ & $-$\tabularnewline
\hline 
$0\left(\frac{1}{2}^{-}\right)$ & \multirow{6}{*}{$\Xi_{c}^{\ast}\bar{D}^{\ast}~[4653.7]$} & $\left[4649.8_{-4.0}^{+3.0}\right]_{B}$ & $J/\psi\Lambda,\eta_{c}\Lambda,\bar{D}_{s}^{(\ast)}\Lambda_{c}$ & \multirow{6}{*}{$\Xi_{c}^{\ast}D^{\ast}$} & $-$ & $-$\tabularnewline
$0\left(\frac{3}{2}^{-}\right)$ &  & $\left[4650.7_{-2.3}^{+1.8}\right]_{B}$ & $J/\psi\Lambda,\bar{D}_{s}^{\ast}\Lambda_{c}$ &  & $-$ & $-$\tabularnewline
$0\left(\frac{5}{2}^{-}\right)$ &  & $\left[4652.0_{-2.4}^{+1.5}\right]_{B}$ & $J/\psi\Lambda,\eta_{c}\Lambda,\bar{D}_{s}^{(\ast)}\Lambda_{c}$ &  & $\left[4645.7,4646.1\right]_{B}$ & $\Xi_{c}D,\Xi_{cc}^{(\ast)}\bar{K},\Omega_{cc}^{(\ast)}\omega$\tabularnewline
$1\left(\frac{1}{2}^{-}\right)$ &  & $-$ & $-$ &  & $\left[4646.6,4653.7\right]_{B}^{\dagger}$ & $\Xi_{c}D,\Xi_{cc}\bar{K},\Omega_{cc}\pi$\tabularnewline
$1\left(\frac{3}{2}^{-}\right)$ &  & $-$ & $-$ &  & $\left[4459.3,4648.1\right]_{V}$ & $\Xi_{cc}^{\ast}\bar{K},\Omega_{cc}^{\ast}\pi,\Omega_{cc}\rho$\tabularnewline
$1\left(\frac{5}{2}^{-}\right)$ &  & $-$ & $-$ &  & $-$ & $-$\tabularnewline
\bottomrule[0.8pt]
\end{tabular}
}
\end{table*}

\subsection{$\Sigma_c^{(\ast)}K^\ast$, $\Xi_c^{(\prime,\ast)}K^\ast$ and $\Sigma_c^{(\ast)}\bar{K}^\ast$, $\Xi_c^{(\prime,\ast)}\bar{K}^\ast$ systems}

The recently observed $T_{cs0}(2900)$ and $T_{c\bar{s}0}^a(2900)$ states have emerged as compelling candidates for the $\bar{D}^\ast K^\ast$ and $D^\ast K^\ast$ molecules, respectively, as evidenced in Refs.~\cite{Wang:2023hpp,Wang:2021lwy} and the references therein. This leads to an intriguing exploration of potential molecular states in systems such as $\Sigma_c^{(\ast)}K^\ast$, $\Xi_c^{(\prime,\ast)}K^\ast$, $\Sigma_c^{(\ast)}\bar{K}^\ast$, and $\Xi_c^{(\prime,\ast)}\bar{K}^\ast$. Very recently, Chen \etal~obtained two molecular states with $\frac{1}{2}(\frac{1}{2}^-)$ and $\frac{1}{2}(\frac{3}{2}^-)$ in the coupled $\Lambda_c K^\ast$$-$$\Sigma_c K^\ast$ system through the OBE model~\cite{Chen:2023qlx}. Additionally, Liu \etal~predicted two isospin-$\frac{1}{2}$ resonances in the $\Sigma_c K^\ast$ system employing a chiral quark model~\cite{Liu:2023oyc}.

In the context of \heavy quark symmetry, these hadronic molecules are expected to be counterparts to the aforementioned states discussed in Sec.~\ref{sec:specSD}, as depicted in Fig.~\ref{fig:correspondences}. The mass spectra and selected decay channels for these systems are presented in Table~\ref{tab:specSK}, while an overview of the spectrum can be found in the third row of Fig.~\ref{fig:spectrum1}.

Within our framework, the effective potentials for $\Sigma_c^{(\ast)}K^\ast$ systems are analogous to those for $\Sigma_c^{(\ast)}\bar{D}^\ast$. Consequently, the mass spectra of these systems exhibit striking similarities. Notably, the molecular states in $\Sigma_c^{(\ast)}K^\ast$ survive in the isospin-$\frac{1}{2}$ channels, just as their counterparts, and are likely to exist as near-threshold virtual (bound) states due to their lower reduced masses. The $\frac{1}{2}(\frac{1}{2}^-)$ and $\frac{1}{2}(\frac{3}{2}^-)$ states in $\Sigma_c K^\ast$ can be identified as charm-strange partners to $P_{\psi}^N(4440)$ and $P_{\psi}^N(4457)$, with their $D_s N$ and $D_s^\ast N$ decay channels corresponding to the $\eta_c N$ and $J/\psi N$ decay channels of the $P_{\psi}^N$ states, respectively. It is reasonable to expect that a similar broad peak, akin to those observed for $T_{cs0}(2900)$ and $T_{c\bar{s}0}^a(2900)$, may manifest in the $D_s^{(\ast)}N$ invariant mass distributions near the $\Sigma_c^{(\ast)}K^\ast$ thresholds.

In the $\Xi_c^{(\prime,\ast)}K^\ast$ systems, we predict the presence of near-threshold virtual states in the isoscalar channels. Serving as the counterparts to the $\Xi_c^{(\prime,\ast)}\bar{D}^\ast$ states, the spectrum of the $\Xi_c^{(\prime,\ast)}K^\ast$ systems mirrors that of the $P_{\psi s}^\Lambda$ states outlined in Table~\ref{tab:specSD} and Fig.~\ref{fig:spectrum1}. Specifically, the $0(\frac{1}{2}^-)$ and $0(\frac{3}{2}^-)$ states in $\Xi_c^{(\prime)}K^\ast$, along with the $0(\frac{1}{2}^-)$, $0(\frac{3}{2}^-)$, and $0(\frac{5}{2}^-)$ states in $\Xi_c^{\ast}K^\ast$, emerge as the promising candidates of the molecular states.

In the $\Sigma_c^{(\ast)}\bar{K}^\ast$ systems, five molecular states are predicted: the $\frac{1}{2}(\frac{3}{2}^-)$ and $\frac{3}{2}(\frac{1}{2}^-)$ states in $\Sigma_c\bar{K}^\ast$, as well as the $\frac{1}{2}(\frac{5}{2}^-)$, $\frac{3}{2}(\frac{1}{2}^-)$, and $\frac{3}{2}(\frac{3}{2}^-)$ states in $\Sigma_c^\ast\bar{K}^\ast$.

Lastly, in the $\Xi_c^{(\prime,\ast)}\bar{K}^\ast$ systems, the $0(\frac{3}{2}^-)$ and $1(\frac{1}{2}^-)$ states in $\Xi_c\bar{K}^\ast$, the $0(\frac{1}{2}^-)$ state in $\Xi_c^\prime\bar{K}^\ast$, and the $0(\frac{5}{2}^-)$ and $1(\frac{1}{2}^-)$ states in $\Xi_c^\ast\bar{K}^\ast$ are all promising candidates of the molecular states. See also the coupled-channel calculations within the OBE model in Ref.~\cite{Chen:2017xat}.

\begin{table*}[!ht]
\centering
\renewcommand{\arraystretch}{1.4}
\caption{The mass spectra and selected decay channels of the molecular pentaquarks in $\Sigma_c^{(\ast)}K^\ast$, $\Xi_c^{(\prime,\ast)}K^\ast$ and $\Sigma_c^{(\ast)}\bar{K}^\ast$, $\Xi_c^{(\prime,\ast)}\bar{K}^\ast$ systems. The notations are the same as those in Table~\ref{tab:specSD}.\label{tab:specSK}}
\setlength{\tabcolsep}{1.15mm}
{
\begin{tabular}{ccccccc}
\toprule[0.8pt]
$I(J^{P})$ & Systems {[}th.{]} & $E_{B}/E_{V}$ & Decay channels & Systems & $E_{B}/E_{V}$ & Decay channels\tabularnewline
\hline 
$\frac{1}{2}\left(\frac{1}{2}^{-}\right)$ & \multirow{4}{*}{$\Sigma_{c}K^{\ast}$ $[3345.5]$} & $\left[3345.4_{-1.8}^{+0.1}\right]_{V}^{\dagger}$ & $D_{s}^{(\ast)}N,\Lambda_{c}K$ & \multirow{4}{*}{$\Sigma_{c}\bar{K}^{\ast}$} & $-$ & $-$\tabularnewline
$\frac{1}{2}\left(\frac{3}{2}^{-}\right)$ &  & $\left[3344.8_{-2.5}^{+0.6}\right]_{V}^{\dagger}$ & $D_{s}^{\ast}N,\Lambda_{c}K^{\ast}$ &  & $\left[3342.4,3345.0\right]_{V}$ & $D^{\ast}\Lambda(\Sigma),\Xi_{c}\rho(\omega),\Xi_{c}^{\ast}\pi,\Lambda_{c}(\Sigma_{c})\bar{K}\pi$\tabularnewline
$\frac{3}{2}\left(\frac{1}{2}^{-}\right)$ &  & $-$ & $-$ &  & $\left[3337.0,3345.5\right]_{B}^{\dagger}$ & $D^{(\ast)}\Sigma,\Xi_{c}\pi,\Xi_{c}\rho,\Sigma_{c}\bar{K}\pi$\tabularnewline
$\frac{3}{2}\left(\frac{3}{2}^{-}\right)$ &  & $-$ & $-$ &  & $-$ & $-$\tabularnewline
\hline 
$\frac{1}{2}\left(\frac{1}{2}^{-}\right)$ & \multirow{6}{*}{$\Sigma_{c}^{\ast}K^{\ast}$ $[3410.1]$} & $\left[3409.5_{-1.2}^{+0.6}\right]_{V}^{\dagger}$ & $D_{s}^{(\ast)}N,\Lambda_{c}K$ & \multirow{6}{*}{$\Sigma_{c}^{\ast}\bar{K}^{\ast}$} & $-$ & $-$\tabularnewline
$\frac{1}{2}\left(\frac{3}{2}^{-}\right)$ &  & $\left[3409.2_{-1.0}^{+0.3}\right]_{V}^{\dagger}$ & $D_{s}^{\ast}N,\Lambda_{c}K^{\ast}$ &  & $-$ & $-$\tabularnewline
$\frac{1}{2}\left(\frac{5}{2}^{-}\right)$ &  & $\left[3409.0_{-3.6}^{+0.7}\right]_{V}^{\dagger}$ & $D_{s}^{(\ast)}N,\Lambda_{c}K^{(\ast)}$ &  & $\left[3408.6,3408.8\right]_{B}$ & $D^{(\ast)}\Lambda(\Sigma),\Xi_{c}\rho(\omega),\Xi_{c}^{(\ast)}\pi,\Lambda_{c}(\Sigma_{c})\bar{K}$\tabularnewline
$\frac{3}{2}\left(\frac{1}{2}^{-}\right)$ &  & $-$ & $-$ &  & $\left[3390.0,3409.9\right]_{B}$ & $D^{(\ast)}\Sigma,\Xi_{c}\pi,\Xi_{c}\rho,\Sigma_{c}\bar{K}$\tabularnewline
$\frac{3}{2}\left(\frac{3}{2}^{-}\right)$ &  & $-$ & $-$ &  & $\left[3161.7,3406.2\right]_{V}$ & $D^{\ast}\Sigma,\Xi_{c}^{\ast}\pi,\Xi_{c}\rho,\Sigma_{c}\bar{K}\pi$\tabularnewline
$\frac{3}{2}\left(\frac{5}{2}^{-}\right)$ &  & $-$ & $-$ &  & $-$ & $-$\tabularnewline
\hline 
$0\left(\frac{1}{2}^{-}\right)$ & \multirow{4}{*}{$\Xi_{c}K^{\ast}$ $[3361.1]$} & $\left[3359.7_{-6.6}^{+1.3}\right]_{V}$ & $D_{s}^{(\ast)}\Lambda,\Lambda_{c}\phi(\eta),\Xi_{c}K\pi$ & \multirow{4}{*}{$\Xi_{c}\bar{K}^{\ast}$} & $-$ & $-$\tabularnewline
$0\left(\frac{3}{2}^{-}\right)$ &  & $\left[3356.0_{-7.6}^{+3.7}\right]_{V}^{\dagger}$ & $D_{s}^{\ast}\Lambda,\Lambda_{c}\phi,\Xi_{c}K\pi$ &  & $\left[3361.0,3361.1\right]_{V}$ & $D^{\ast}\Xi,\Xi_{c}\bar{K}\pi,\Omega_{c}\pi^{+}\pi^{-}\pi^{0}$\tabularnewline
$1\left(\frac{1}{2}^{-}\right)$ &  & $-$ & $-$ &  & $\left[3351.9,3361.1\right]_{V}^{\dagger}$ & $D^{(\ast)}\Xi,\Xi_{c}\bar{K}\pi,\Omega_{c}\pi\pi$\tabularnewline
$1\left(\frac{3}{2}^{-}\right)$ &  & $-$ & $-$ &  & $-$ & $-$\tabularnewline
\hline 
$0\left(\frac{1}{2}^{-}\right)$ & \multirow{4}{*}{$\Xi_{c}^{\prime}K^{\ast}$ $[3470.2]$} & $\left[3466.1_{-5.3}^{+2.8}\right]_{V}$ & $D_{s}^{(\ast)}\Lambda,\Lambda_{c}\phi(\eta),\Xi_{c}K$ & \multirow{4}{*}{$\Xi_{c}^{\prime}\bar{K}^{\ast}$} & $\left[3460.9,3466.4\right]_{V}$ & $D^{(\ast)}\Xi,\Xi_{c}\bar{K},\Omega_{c}\pi^{+}\pi^{-}\pi^{0}$\tabularnewline
$0\left(\frac{3}{2}^{-}\right)$ &  & $\left[3467.5_{-3.1}^{+1.8}\right]_{V}$ & $D_{s}^{\ast}\Lambda,\Lambda_{c}\phi,\Xi_{c}K\pi$ &  & $-$ & $-$\tabularnewline
$1\left(\frac{1}{2}^{-}\right)$ &  & $-$ & $-$ &  & $-$ & $-$\tabularnewline
$1\left(\frac{3}{2}^{-}\right)$ &  & $-$ & $-$ &  & $\left[\sim3134.1\right]_{V}^{\sharp}$ & $D^{\ast}\Xi,\Xi_{c}\bar{K}\pi,\Omega_{c}\pi\pi$\tabularnewline
\hline 
$0\left(\frac{1}{2}^{-}\right)$ & \multirow{6}{*}{$\Xi_{c}^{\ast}K^{\ast}$ $[3537.1]$} & $\left[3535.8_{-5.5}^{+1.3}\right]_{V}$ & $D_{s}^{(\ast)}\Lambda,\Lambda_{c}\phi(\eta),\Xi_{c}K$ & \multirow{6}{*}{$\Xi_{c}^{\ast}\bar{K}^{\ast}$} & $-$ & $-$\tabularnewline
$0\left(\frac{3}{2}^{-}\right)$ &  & $\left[3534.9_{-3.6}^{+1.7}\right]_{V}$ & $D_{s}^{\ast}\Lambda,\Lambda_{c}\phi,\Xi_{c}K\pi$ &  & $-$ & $-$\tabularnewline
$0\left(\frac{5}{2}^{-}\right)$ &  & $\left[3532.7_{-7.1}^{+3.3}\right]_{V}$ & $D_{s}^{(\ast)}\Lambda,\Lambda_{c}\phi(\eta),\Xi_{c}K$ &  & $\left[3536.5,3536.8\right]_{B}^{\dagger}$ & $D^{(\ast)}\Xi,\Xi_{c}\bar{K},\Omega_{c}\omega$\tabularnewline
$1\left(\frac{1}{2}^{-}\right)$ &  & $-$ & $-$ &  & $\left[3514.2,3537.1\right]_{V}$ & $D^{(\ast)}\Xi,\Xi_{c}\bar{K},\Omega_{c}\pi$\tabularnewline
$1\left(\frac{3}{2}^{-}\right)$ &  & $-$ & $-$ &  & $\left[\sim3473.8\right]_{V}^{\sharp}$ & $D^{\ast}\Xi,\Xi_{c}\bar{K}\pi,\Omega_{c}^{\ast}\pi,\Omega_{c}\rho$\tabularnewline
$1\left(\frac{5}{2}^{-}\right)$ &  & $-$ & $-$ &  & $-$ & $-$\tabularnewline
\bottomrule[0.8pt]
\end{tabular}
}
\end{table*}

\subsection{$\Xi_{cc}^{(\ast)}\bar{D}^{(\ast)}$, $\Xi_{cc}^{(\ast)}K^{\ast}$ and $\Xi_{cc}^{(\ast)}D^{(\ast)}$, $\Xi_{cc}^{(\ast)}\bar{K}^{\ast}$ systems}

The molecular pentaquark systems involving a double-charm baryon have attracted significant interest due to the presence of the heavy diquark-antiquark symmetry (HDAS)~\cite{Meng:2022ozq,Savage:1990di,Fleming:2005pd,Meng:2018zbl}. According to HDAS, the double-heavy baryons $\Xi_{cc}^{(\ast)}$ can be correlated with the anti-charmed mesons $\bar{D}^{(\ast)}$. Consequently, the existence of hadronic molecules in the $DD^\ast$ system---the $T_{cc}(3875)$, and the $D\bar{D}^\ast$ system---the $X(3872)$ and $Z_c(3900)$, implies the potential presence of pentaquarks in the $\Xi_{cc}^{(\ast)}\bar{D}^{(\ast)}$ and $\Xi_{cc}^{(\ast)}D^{(\ast)}$ systems, respectively. Previous studies have predicted several molecular states in these systems. For example, Guo \etal~predicted a bound state with $0(\frac{5}{2}^-)$ in the $\Xi_{cc}^\ast D^\ast$ system by utilizing the $X(3872)$ as an input~\cite{Guo:2013xga}. Additionally, Chen \etal~derived bound states with $0(\frac{1}{2}^-)$ in the $\Xi_{cc}\bar{D}^\ast$ system, $0(\frac{1}{2}^-)$ in the $\Xi_{cc}D$ system, and $0(\frac{3}{2}^-)$ in the $\Xi_{cc}D^\ast$ system~\cite{Chen:2017jjn}. Recently, Asanuma \etal~proposed a connection between $T_{cc}$ and the molecules in $\Xi_{cc}^{(\ast)}D^{(\ast)}$, and they predicted a bound state with $0(\frac{1}{2}^-)$~\cite{Asanuma:2023atv}. The investigation of triple-heavy pentaquarks has also been pursued using various approaches such as the QCD sum rule~\cite{Wang:2018ihk}, the chromomagnetic interaction model~\cite{Li:2018vhp,An:2019idk}, the OBE model~\cite{Liu:2018bkx,Wang:2019aoc}, and the quark model~\cite{Yan:2023iie}.

\begin{figure*}[!ht]
\begin{centering}
    \scalebox{1.0}{\includegraphics[width=\linewidth]{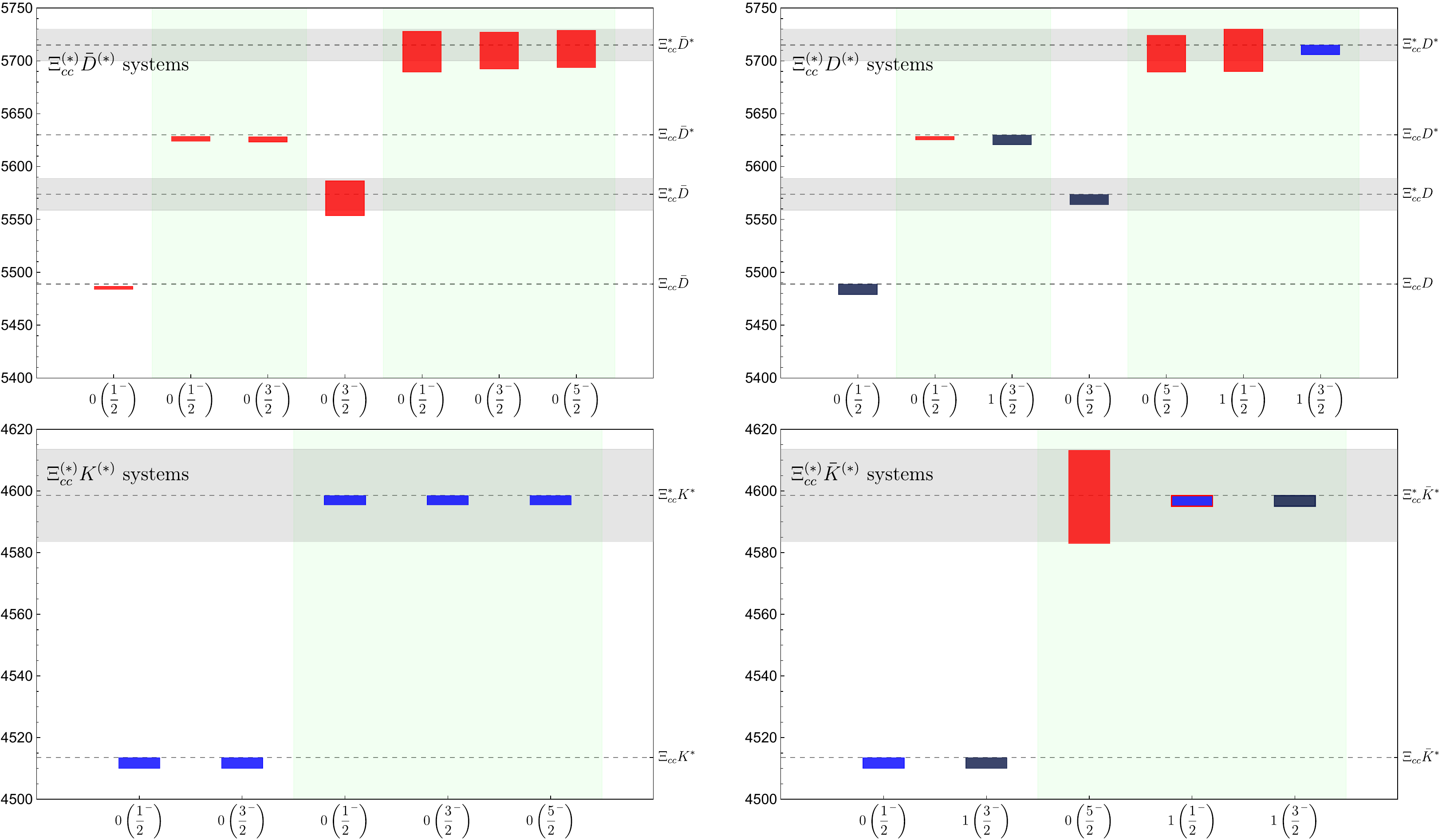}}
    \caption{Spectrum of the molecular pentaquarks with a double-charm baryon. The horizontal bands with light gray represent the uncertainties of $\Xi_{cc}^\ast$ mass. Other notations are the same as those in Fig.~\ref{fig:spectrum1}.\label{fig:spectrum2}}
\end{centering}
\end{figure*}

In this study, we extend our analysis to include pentaquark systems with double-heavy baryons beyond the triple-charm $\Xi_{cc}^{(\ast)}\bar{D}^{(\ast)}$ and $\Xi_{cc}^{(\ast)}D^{(\ast)}$ systems. We also consider the $\Xi_{cc}^{(\ast)}K^{\ast}$ and $\Xi_{cc}^{(\ast)}\bar{K}^{\ast}$ systems. The numerical results for these systems are presented in Table~\ref{tab:specXDK}, and the corresponding spectrum is illustrated in Fig.~\ref{fig:spectrum2}. It is important to note that the mass of the $\Xi_{cc}^{++}$ has been measured to be approximately $3621.5$ MeV~\cite{Workman:2022ynf}, while the $\Xi_{cc}^\ast$ is unobserved yet. Within the framework of HDAS, the mass splitting $m_{\Xi_{cc}^\ast}-m_{\Xi_{cc}}= \frac{3}{4}(m_{D^\ast}-m_D)\simeq 106$ MeV~\cite{Savage:1990di,Fleming:2005pd}. Quark model calculations, such as in Ref.~\cite{Lu:2017meb}, predicted the $m_{\Xi_{cc}^\ast}-m_{\Xi_{cc}}$ to be around $70$ MeV. In our calculations, we consider $m_{\Xi_{cc}^\ast}-m_{\Xi_{cc}}$ values in the range of $[70,100]$ MeV to cover the estimates from HDAS and quark models. From Table~\ref{tab:specXDK}, it can be observed that the notation $\pm15$ MeV is used to represent the uncertainties associated with the mass of $\Xi_{cc}^\ast$.

In our framework, the connections between the $DD^\ast$ and $\Xi_{cc}^{(\ast)}\bar{D}^{(\ast)}$ systems are presented in a more direct manner through their interacting types, i.e., the $V_{qq}$ interactions. Thus, the existence of the molecule candidate $T_{cc}$ in the $DD^\ast$ system naturally leads to the existence of molecules in the $\Xi_{cc}^{(\ast)}\bar{D}^{(\ast)}$ systems. The mass spectrum of the $\Xi_{cc}^{(\ast)}\bar{D}^{(\ast)}$ systems exhibits similarities to that of the $\Xi_{c}^{\prime,\ast}\bar{D}^{(\ast)}$ systems. For instance, all seven bound states exist in the isoscalar channels, while there are no molecules in the isovector channels. In addition to the $0(\frac{1}{2}^-)$ states reported in Refs.~\cite{Chen:2017jjn,Asanuma:2023atv}, we also obtain $0(\frac{3}{2}^-)$ and $0(\frac{5}{2}^-)$ states. The seven molecules with $0(\frac{1}{2}^-)$ in $\Xi_{cc}\bar{D}$, $0(\frac{1}{2}^-)$ and $0(\frac{3}{2}^-)$ in $\Xi_{cc}\bar{D}^\ast$, $0(\frac{3}{2}^-)$ in $\Xi_{cc}^\ast\bar{D}$, and $0(\frac{1}{2}^-)$, $0(\frac{3}{2}^-)$, and $0(\frac{5}{2}^-)$ in $\Xi_{cc}^\ast\bar{D}^\ast$ are good candidates of the molecular states. These states can be reconstructed, for example, from the $J/\psi\Lambda_c$ invariant mass distributions, and they can be regarded as analogues of the $J/\psi\Lambda$ channel for the $P_{\psi s}^\Lambda$ states. The mass spectrum in the $\Xi_{cc}^{(\ast)}K^\ast$ systems follows a similar pattern, although the five isoscalar molecules become virtual states due to their low reduced masses.

Regarding the $\Xi_{cc}^{(\ast)}D^{(\ast)}$ systems, we also find a $0(\frac{5}{2}^-)$ state in $\Xi_{cc}^\ast D^\ast$ with a binding energy of approximately $10$ MeV\footnote{One can easily verify the potential of $\Xi_{cc}^\ast D^\ast$ with $0(\frac{5}{2}^-)$ is the same as that of $D\bar{D}^\ast$ with $0^+(1^{++})$ [where the $X(3872)$ exists], and they both equal to $-\frac{5}{2}\tilde{c}_{s}-\frac{5}{2}\tilde{c}_{a}$. See Table~\ref{tab:potentials}, Fig.~\ref{fig:correspondences} and Ref.~\cite{Wang:2023hpp}.}. The prediction in Ref.~\cite{Guo:2013xga} using the HDAS is consistent with our calculations. Additionally, we identify two more bound states: one with $0(\frac{1}{2}^-)$ in $\Xi_{cc}D^\ast$ and another with $1(\frac{1}{2}^-)$ in $\Xi_{cc}^\ast D^\ast$. We also find one virtual state with $1(\frac{3}{2}^-)$ in $\Xi_{cc}^\ast D^\ast$. The molecule spectrum of the $\Xi_{cc}^{(\ast)}\bar{K}^\ast$ system is similar to that of the $\Xi_{cc}^{(\ast)}D^\ast$ system. Further details can be found in Table~\ref{tab:specXDK} and Fig.~\ref{fig:spectrum2}.

\begin{table*}[!ht]
\centering
\renewcommand{\arraystretch}{1.5}
\caption{The mass spectra and selected decay channels of the molecular pentaquarks in $\Xi_{cc}^{(\ast)}\bar{D}^{(\ast)}$, $\Xi_{cc}^{(\ast)}K^\ast$ and $\Xi_{cc}^{(\ast)}D^{(\ast)}$, $\Xi_{cc}^{(\ast)}\bar{K}^\ast$ systems. The notations are the same as those in Table~\ref{tab:specSD}.\label{tab:specXDK}}
\setlength{\tabcolsep}{0.3mm}
{
\begin{tabular}{ccccccc}
\toprule[0.8pt]
$I(J^{P})$ & Systems {[}th.{]} & $E_{B}/E_{V}$ & Decay channels & Systems & $E_{B}/E_{V}$ & Decay channels\tabularnewline
\hline 
$0\left(\frac{1}{2}^{-}\right)$ & \multirow{2}{*}{$\Xi_{cc}\bar{D}$ $[5488.8]$} & $\left[5485.3_{-1.4}^{+1.3}\right]_{B}$ & $J/\psi\Lambda_{c},\eta_{c}\Lambda_{c}$ & \multirow{2}{*}{$\Xi_{cc}D$} & $\left[\sim5476.9\right]_{V}^{\sharp}$ & $\Omega_{ccc}\pi^{+}\pi^{-}\pi^{0}$\tabularnewline
$1\left(\frac{1}{2}^{-}\right)$ &  & $-$ & $-$ &  & $-$ & $-$\tabularnewline
\hline 
$0\left(\frac{1}{2}^{-}\right)$ & \multirow{4}{*}{$\Xi_{cc}\bar{D}^{\ast}$ $[5630.1]$} & $\left[5626.5_{-2.5}^{+2.0}\right]_{B}$ & $J/\psi\Lambda_{c},\eta_{c}\Lambda_{c}$ & \multirow{4}{*}{$\Xi_{cc}D^{\ast}$} & $\left[5626.3,5628.5\right]_{B}$ & $\Xi_{cc}D\pi,\Omega_{ccc}\pi^{+}\pi^{-}\pi^{0}$\tabularnewline
$0\left(\frac{3}{2}^{-}\right)$ &  & $\left[5625.5_{-2.1}^{+1.8}\right]_{B}$ & $J/\psi\Lambda_{c}$ &  & $-$ & $-$\tabularnewline
$1\left(\frac{1}{2}^{-}\right)$ &  & $-$ & $-$ &  & $-$ & $-$\tabularnewline
$1\left(\frac{3}{2}^{-}\right)$ &  & $-$ & $-$ &  & $\left[\sim5591.4\right]_{V}^{\sharp}$ & $\Xi_{cc}D\pi,\Omega_{ccc}\pi\pi$\tabularnewline
\hline 
$0\left(\frac{3}{2}^{-}\right)$ & \multirow{2}{*}{$\Xi_{cc}^{\ast}\bar{D}$ $[5573.8\pm15]$} & $\left[5570.2_{-1.5}^{+1.3}\pm15\right]_{B}$ & $J/\psi\Lambda_{c}$ & \multirow{2}{*}{$\Xi_{cc}^{\ast}D$} & $\left[\sim5562.3\pm15\right]_{V}^{\sharp}$ & $\Omega_{ccc}\pi^{+}\pi^{-}\pi^{0}$\tabularnewline
$1\left(\frac{3}{2}^{-}\right)$ &  & $-$ & $-$ &  & $-$ & $-$\tabularnewline
\hline 
$0\left(\frac{1}{2}^{-}\right)$ & \multirow{6}{*}{$\Xi_{cc}^{\ast}\bar{D}^{\ast}$$[5715.1\pm15]$} & $\left[5709.0_{-4.6}^{+3.8}\pm15\right]_{B}$ & $J/\psi\Lambda_{c},\eta_{c}\Lambda_{c},\Xi_{cc}\bar{D}$ & \multirow{6}{*}{$\Xi_{cc}^{\ast}D^{\ast}$} & $-$ & $-$\tabularnewline
$0\left(\frac{3}{2}^{-}\right)$ &  & $\left[5710.0_{-2.7}^{+2.3}\pm15\right]_{B}$ & $J/\psi\Lambda_{c},\Xi_{cc}\bar{D}^{\ast},\Xi_{cc}^{\ast}\bar{D}$ &  & $-$ & $-$\tabularnewline
$0\left(\frac{5}{2}^{-}\right)$ &  & $\left[5711.7_{-3.0}^{+2.3}\pm15\right]_{B}$ & $J/\psi\Lambda_{c},\eta_{c}\Lambda_{c},\Xi_{cc}\bar{D}$ &  & $\left[5704.6\pm0.2\pm15\right]_{B}$ & $\Xi_{cc}D,\Omega_{ccc}\omega$\tabularnewline
$1\left(\frac{1}{2}^{-}\right)$ &  & $-$ & $-$ &  & $\left[5710.0\pm5.0\pm15\right]_{B}$ & $\Xi_{cc}D,\Omega_{ccc}\pi,\Omega_{ccc}\rho$\tabularnewline
$1\left(\frac{3}{2}^{-}\right)$ &  & $-$ & $-$ &  & $\left[5649.7\pm63.3\pm15\right]_{V}$ & $\Xi_{cc}D\pi,\Omega_{ccc}\rho$\tabularnewline
$1\left(\frac{5}{2}^{-}\right)$ &  & $-$ & $-$ &  & $-$ & $-$\tabularnewline
\hline 
$0\left(\frac{1}{2}^{-}\right)$ & \multirow{4}{*}{$\Xi_{cc}K^{\ast}$ $[4513.6]$} & $\left[4511.6_{-3.6}^{+1.6}\right]_{V}$ & $D_{s}^{(\ast)}\Lambda_{c},\Xi_{cc}K,\Xi_{cc}K\pi$ & \multirow{4}{*}{$\Xi_{cc}\bar{K}^{\ast}$} & $\left[4508.0,4511.8\right]_{V}$ & $D^{(\ast)}\Xi_{c},\Xi_{cc}\bar{K}\pi,\Omega_{cc}\pi^{+}\pi^{-}\pi^{0}$\tabularnewline
$0\left(\frac{3}{2}^{-}\right)$ &  & $\left[4512.5_{-2.0}^{+0.9}\right]_{V}$ & $D_{s}^{\ast}\Lambda_{c},\Xi_{cc}K\pi$ &  & $-$ & $-$\tabularnewline
$1\left(\frac{1}{2}^{-}\right)$ &  & $-$ & $-$ &  & $-$ & $-$\tabularnewline
$1\left(\frac{3}{2}^{-}\right)$ &  & $-$ & $-$ &  & $\left[\sim4242.6\right]_{V}^{\sharp}$ & $D^{\ast}\Xi_{c},\Xi_{cc}\bar{K}\pi,\Omega_{cc}^{\ast}\pi,\Omega_{cc}\pi\pi$\tabularnewline
\hline 
$0\left(\frac{1}{2}^{-}\right)$ & \multirow{6}{*}{$\Xi_{cc}^{\ast}K^{\ast}$ $[4598.6\pm15]$} & $\left[4598.2_{-3.5}^{+0.3}\pm15\right]_{V}$ & $D_{s}^{(\ast)}\Lambda_{c},\Xi_{cc}K$ & \multirow{6}{*}{$\Xi_{cc}^{\ast}\bar{K}^{\ast}$} & $-$ & $-$\tabularnewline
$0\left(\frac{3}{2}^{-}\right)$ &  & $\left[4597.7_{-2.4}^{+0.8}\pm15\right]_{V}$ & $D_{s}^{\ast}\Lambda_{c},\Xi_{cc}K\pi$ &  & $-$ & $-$\tabularnewline
$0\left(\frac{5}{2}^{-}\right)$ &  & $\left[4596.4_{-5.0}^{+2.0}\pm15\right]_{V}$ & $D_{s}^{(\ast)}\Lambda_{c},\Xi_{cc}K$ &  & $\left[4598.1\pm0.1\pm15\right]_{B}$ & $D^{(\ast)}\Xi_{c},\Xi_{cc}\bar{K},\Omega_{cc}\omega$\tabularnewline
$1\left(\frac{1}{2}^{-}\right)$ &  & $-$ & $-$ &  & $\left[4590.6\pm8.0\pm15\right]_{V}^{\dagger}$ & $D^{(\ast)}\Xi_{c},\Xi_{cc}\bar{K},\Omega_{cc}\pi,\Omega_{cc}\rho$\tabularnewline
$1\left(\frac{3}{2}^{-}\right)$ &  & $-$ & $-$ &  & $\left[\sim4550.7\pm15\right]_{V}^{\sharp}$ & $D^{\ast}\Xi_{c},\Xi_{cc}\bar{K}\pi,\Omega_{cc}^{\ast}\pi,\Omega_{cc}\rho$\tabularnewline
$1\left(\frac{5}{2}^{-}\right)$ &  & $-$ & $-$ &  & $-$ & $-$\tabularnewline
\bottomrule[0.8pt]
\end{tabular}
}
\end{table*}

\subsection{$\Sigma^{(\ast)}\bar{D}^{(\ast)}$, $\Xi^{(\ast)}\bar{D}^{(\ast)}$ and $\Sigma^{(\ast)}D^{(\ast)}$, $\Xi^{(\ast)}D^{(\ast)}$ systems}

In recent years, there have been significant advancements in the study of pentaquark states with different quark configurations, particularly in the field of hadronic molecules. One prominent example is the observation of pentaquarks with the configurations $\bar{c}cqqq$ and $\bar{c}csqq$, known as the $P_{\psi}^N$ and $P_{\psi s}^\Lambda$, respectively, which have been confirmed through experiments conducted by LHCb. Prior to the discovery of hidden-charm pentaquarks, Gignoux \etal~\cite{Gignoux:1987cn} and Lipkin~\cite{Lipkin:1987sk} had already investigated anti-charm strange pentaquarks with the configuration $\bar{c}sqqq$. It was denoted as the $P_{\bar{c}s}$ in Ref.~\cite{Lipkin:1987sk} and it corresponds to a bound state of the $\bar{D}_s N$ system, despite that the residual strong interaction between $\bar{D}_s$ and $N$ should be too weak to form a bound state. Later, Hofmann \etal~predicted the existence of isospin-$\frac{1}{2}$ bound states in the coupled systems of $\bar{D}_sN$-$\bar{D}\Lambda$-$\bar{D}\Sigma$, as well as an isoscalar state in the coupled $\bar{D}_s\Lambda$-$\bar{D}\Xi$ systems~\cite{Hofmann:2005sw}. More recently, Yalikun \etal~obtained three bound states with isospin-$\frac{1}{2}$ in the $\Sigma\bar{D}^{(\ast)}$ systems through the OBE model~\cite{Yalikun:2021dpk}, and they proposed to search for these states in the process $B_s^0\to\bar{n}D_s^- p$. Additionally, Yan \etal~obtained a similar spectrum for the $\Sigma\bar{D}^{(\ast)}$ systems based on the light-meson exchange saturation model~\cite{Yan:2023ttx}.

Moreover, pentaquark systems with the quark configuration $cssq\bar{q}$, such as the $\Xi D^{(\ast)}$, have been extensively studied in recent years to understand some of the excited $\Omega_c$ states observed by the LHCb~\cite{LHCb:2017uwr}. Various theoretical approaches, including the chiral unitary approach~\cite{Montana:2017kjw,Debastiani:2017ewu,Nieves:2017jjx,Yu:2018yxl}, OBE related models~\cite{Wang:2017smo,Yan:2023ttx,Liu:2018bkx}, QCD sum rule~\cite{Xin:2023gkf}, and chiral quark model~\cite{Yan:2023tvl}, have been employed to investigate the properties of these systems.

\begin{figure*}[!ht]
\begin{centering}
    \scalebox{1.0}{\includegraphics[width=\linewidth]{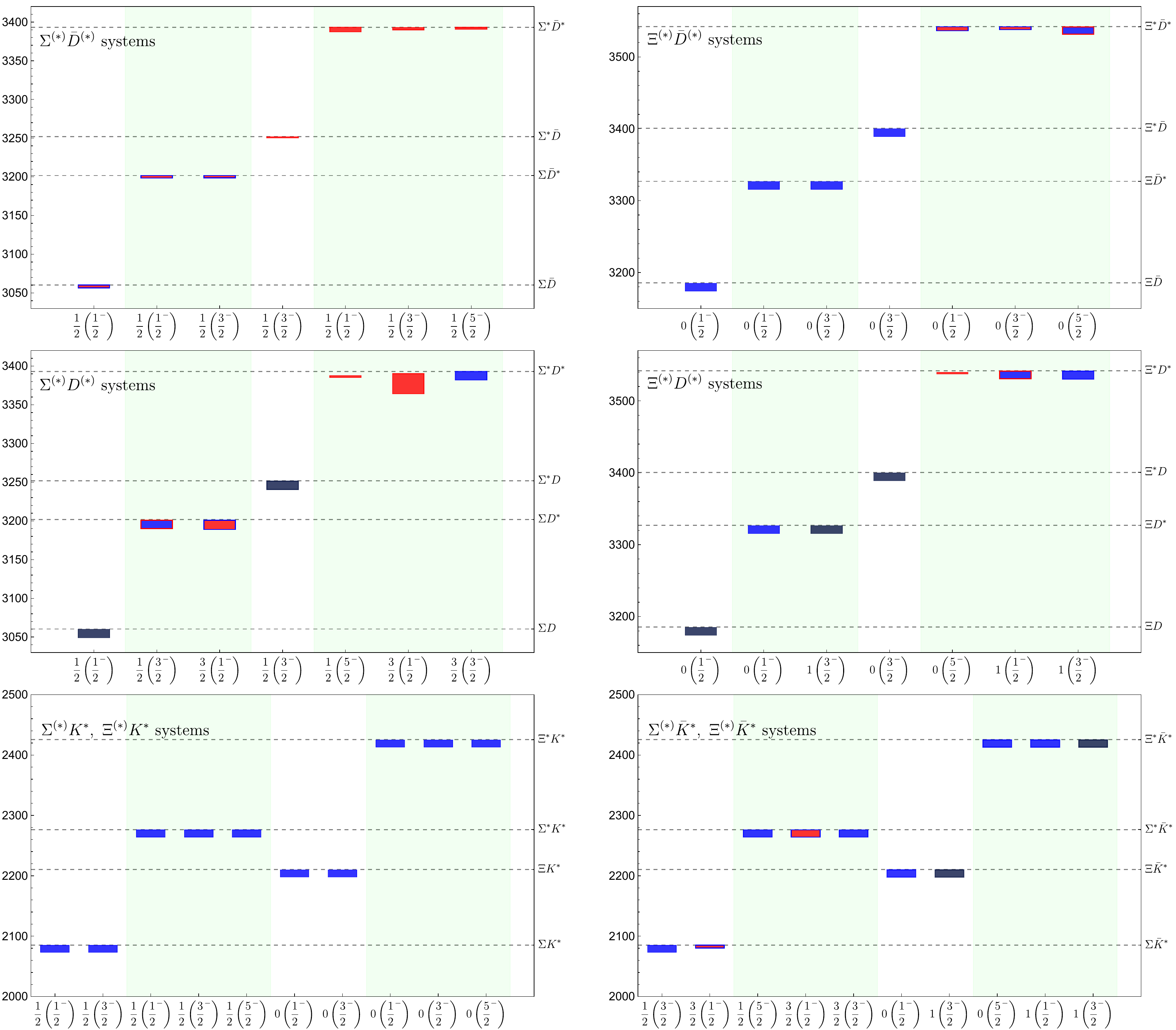}}
    \caption{Spectrum of the molecular pentaquarks with a hyperon $\Sigma^{(\ast)}$ or $\Xi^{(\ast)}$. The corresponding di-hadron systems are marked in each subfigure. The notations are the same as those in Fig.~\ref{fig:spectrum1}.\label{fig:spectrum3}}
\end{centering}
\end{figure*}

In Table~\ref{tab:specHD} and Fig.~\ref{fig:spectrum3}, we present the mass spectra and selected decay channels for the $\Sigma^{(\ast)}\bar{D}^{(\ast)}$, $\Xi^{(\ast)}\bar{D}^{(\ast)}$, $\Sigma^{(\ast)}D^{(\ast)}$, and $\Xi^{(\ast)}D^{(\ast)}$ systems. Our calculation reveals the existence of several bound states and virtual states in these systems. In the $\Sigma^{(\ast)}\bar{D}^{(\ast)}$ systems, we identified seven molecules, all of which are present in the isospin-$\frac{1}{2}$ channels. Among them, the three states in $\Sigma\bar{D}^{(\ast)}$ can be considered near-threshold bound (or virtual) states, while the molecules in $\Sigma^\ast\bar{D}^{(\ast)}$ correspond to shallow bound states. Similarly, we derived seven hadronic molecules in the isoscalar $\Xi^{(\ast)}\bar{D}^{(\ast)}$ systems, with most of them being virtual states. In the $\Sigma^{(\ast)}D^{(\ast)}$ and $\Xi^{(\ast)}D^{(\ast)}$ systems, we identified several states, such as the $\frac{1}{2}(\frac{3}{2}^-)$ and $\frac{3}{2}(\frac{1}{2}^-)$ states in $\Sigma D^\ast$, the $\frac{1}{2}(\frac{5}{2}^-)$, $\frac{3}{2}(\frac{1}{2}^-)$, and $\frac{3}{2}(\frac{3}{2}^-)$ states in $\Sigma^\ast D^\ast$, and the $0(\frac{1}{2}^-)$ state in $\Xi D^\ast$, as well as the $0(\frac{5}{2}^-)$, $1(\frac{1}{2}^-)$, and $1(\frac{3}{2}^-)$ states in $\Xi^\ast D^\ast$, which are strong candidates of the molecular states.

Our calculations indicate that there is no bound state in the $\Xi D$ system, which disfavors the molecular interpretation of the excited $\Omega_c$~\cite{LHCb:2017uwr} states discovered by LHCb. Nevertheless, it is worth mentioning that the LHCb data also hint the presence of a broad structure around $3188$ MeV, close to the $\Xi D$ threshold, which cannot be explained by a simple superposition of other states~\cite{LHCb:2017uwr}. In Ref.~\cite{Wang:2017smo}, this structure was considered to be a $\Xi D$ bound state. In our work, there possibly exists a virtual state with $0(\frac{1}{2}^-)$ in $\Xi D$ (see Table~\ref{tab:specHD}), and it is known that the virtual state pole will give rise to a peak exactly at the $\Xi D$ threshold in the inelastic channels, e.g., the $\Xi_c\bar{K}$. This provides a plausible explanation for this hint from the LHCb data~\cite{LHCb:2017uwr}.

\begin{table*}[!ht]
\centering
\renewcommand{\arraystretch}{1.5}
\caption{The mass spectra and selected decay channels of the molecular pentaquarks in $\Sigma^{(\ast)}\bar{D}^{(\ast)}$, $\Xi^{(\ast)}\bar{D}^{(\ast)}$ and $\Sigma^{(\ast)}D^{(\ast)}$, $\Xi^{(\ast)}D^{(\ast)}$ systems. The notations are the same as those in Table~\ref{tab:specSD}.\label{tab:specHD}}
\setlength{\tabcolsep}{2.9mm}
{
\begin{tabular}{ccccccc}
\toprule[0.8pt]
$I(J^{P})$ & Systems {[}th.{]} & $E_{B}/E_{V}$ & Decay channels & Systems & $E_{B}/E_{V}$ & Decay channels\tabularnewline
\hline 
$\frac{1}{2}\left(\frac{1}{2}^{-}\right)$ & \multirow{2}{*}{$\Sigma\bar{D}$ $[3060.4]$} & $\left[3060.0_{-0.4}^{+0.3}\right]_{B}^{\dagger}$ & $N\bar{D}_{s}^{(\ast)},\Lambda\bar{D}$ & \multirow{2}{*}{$\Sigma D$} & $\left[\sim3002.0\right]_{V}^{\sharp}$ & $\Lambda_{c}\bar{K},\Xi_{c}\pi$\tabularnewline
$\frac{3}{2}\left(\frac{1}{2}^{-}\right)$ &  & $-$ & $-$ &  & $-$ & $-$\tabularnewline
\hline 
$\frac{1}{2}\left(\frac{1}{2}^{-}\right)$ & \multirow{4}{*}{$\Sigma\bar{D}^{\ast}$ $[3201.7]$} & $\left[3201.1_{-2.4}^{+0.6}\right]_{B}^{\dagger}$ & $N\bar{D}_{s}^{(\ast)},\Lambda\bar{D}^{(\ast)}$ & \multirow{4}{*}{$\Sigma D^{\ast}$} & $-$ & $-$\tabularnewline
$\frac{1}{2}\left(\frac{3}{2}^{-}\right)$ &  & $\left[3201.6_{-0.8}^{+0.1}\right]_{B}^{\dagger}$ & $N\bar{D}_{s}^{\ast},\Lambda\bar{D}^{\ast}$ &  & $\left[3201.2,3201.7\right]_{V}^{\dagger}$ & $\Lambda_{c}\bar{K}\pi,\Xi_{c}^{\ast}\pi,\Sigma D\pi$\tabularnewline
$\frac{3}{2}\left(\frac{1}{2}^{-}\right)$ &  & $-$ & $-$ &  & $\left[3188.9,3201.7\right]_{B}^{\dagger}$ & $\Sigma_{c}\bar{K},\Xi_{c}\pi,\Sigma D\pi$\tabularnewline
$\frac{3}{2}\left(\frac{3}{2}^{-}\right)$ &  & $-$ & $-$ &  & $-$ & $-$\tabularnewline
\hline 
$\frac{1}{2}\left(\frac{3}{2}^{-}\right)$ & \multirow{2}{*}{$\Sigma^{\ast}\bar{D}$ $[3251.8]$} & $\left[3251.3_{-0.9}^{+0.5}\right]_{B}$ & $N\bar{D}_{s}^{\ast},\Lambda\bar{D}^{\ast}$ & \multirow{2}{*}{$\Sigma^{\ast}D$} & $\left[\sim3209.8\right]_{V}^{\sharp}$ & $\Lambda_{c}\bar{K}\pi,\Xi_{c}^{\ast}\pi,\Lambda D^{\ast}$\tabularnewline
$\frac{3}{2}\left(\frac{3}{2}^{-}\right)$ &  & $-$ & $-$ &  & $-$ & $-$\tabularnewline
\hline 
$\frac{1}{2}\left(\frac{1}{2}^{-}\right)$ & \multirow{6}{*}{$\Sigma^{\ast}\bar{D}^{\ast}$ $[3393.1]$} & $\left[3391.1_{-3.8}^{+2.0}\right]_{B}$ & $N\bar{D}_{s}^{(\ast)},\Lambda\bar{D}^{(\ast)}$ & \multirow{6}{*}{$\Sigma^{\ast}D^{\ast}$} & $-$ & $-$\tabularnewline
$\frac{1}{2}\left(\frac{3}{2}^{-}\right)$ &  & $\left[3391.9_{-2.0}^{+1.1}\right]_{B}$ & $N\bar{D}_{s}^{\ast},\Lambda\bar{D}^{\ast}$ &  & $-$ & $-$\tabularnewline
$\frac{1}{2}\left(\frac{5}{2}^{-}\right)$ &  & $\left[3392.8_{-1.8}^{+0.4}\right]_{B}$ & $N\bar{D}_{s}^{(\ast)},\Lambda\bar{D}^{(\ast)}$ &  & $\left[3387.3,3387.7\right]_{B}$ & $\Lambda_{c}\bar{K},\Xi_{c}\pi,\Sigma D^{(\ast)},\Lambda D^{(\ast)}$\tabularnewline
$\frac{3}{2}\left(\frac{1}{2}^{-}\right)$ &  & $-$ & $-$ &  & $\left[3364.3,3390.3\right]_{B}$ & $\Sigma_{c}\bar{K},\Xi_{c}\pi,\Sigma D$\tabularnewline
$\frac{3}{2}\left(\frac{3}{2}^{-}\right)$ &  & $-$ & $-$ &  & $\left[3269.9,3393.0\right]_{V}$ & $\Sigma_{c}^{\ast}\bar{K},\Xi_{c}^{\ast}\pi,\Sigma D^{\ast}$\tabularnewline
$\frac{3}{2}\left(\frac{5}{2}^{-}\right)$ &  & $-$ & $-$ &  & $-$ & $-$\tabularnewline
\hline 
$0\left(\frac{1}{2}^{-}\right)$ & \multirow{2}{*}{$\Xi\bar{D}$ $[3185.5]$} & $\left[3185.2_{-0.9}^{+0.3}\right]_{V}$ & $\Lambda\bar{D}_{s}$ & \multirow{2}{*}{$\Xi D$} & $\left[\sim3110.0\right]_{V}$ & $\Xi_{c}\bar{K},\Omega_{c}\pi^{+}\pi^{-}\pi^{0}$\tabularnewline
$1\left(\frac{1}{2}^{-}\right)$ &  & $-$ & $-$ &  & $-$ & $-$\tabularnewline
\hline 
$0\left(\frac{1}{2}^{-}\right)$ & \multirow{4}{*}{$\Xi\bar{D}^{\ast}$ $[3326.8]$} & $[3326.5_{-1.9}^{+0.3}]_{V}$ & $\Lambda\bar{D}_{s}^{(\ast)},\Xi\bar{D}\pi$ & \multirow{4}{*}{$\Xi D^{\ast}$} & $\left[3324.6,3326.5\right]_{V}$ & $\Xi_{c}\bar{K},\Xi D,\Omega_{c}\pi^{+}\pi^{-}\pi^{0}$\tabularnewline
$0\left(\frac{3}{2}^{-}\right)$ &  & $\left[3326.7_{-0.8}^{+0.1}\right]_{V}$ & $\Lambda\bar{D}_{s}^{\ast},\Xi\bar{D}\pi$ &  & $-$ & $-$\tabularnewline
$1\left(\frac{1}{2}^{-}\right)$ &  & $-$ & $-$ &  & $-$ & $-$\tabularnewline
$1\left(\frac{3}{2}^{-}\right)$ &  & $-$ & $-$ &  & $\left[\sim3132.8\right]_{V}$ & $\Xi_{c}^{\ast}\bar{K},\Omega_{c}^{\ast}\pi$\tabularnewline
\hline 
$0\left(\frac{3}{2}^{-}\right)$ & \multirow{2}{*}{$\Xi^{\ast}\bar{D}$ $[3400.7]$} & $\left[3400.0_{-0.3}^{+0.6}\right]_{V}$ & $\Lambda\bar{D}_{s}^{\ast},\Xi\bar{D}\pi$ & \multirow{2}{*}{$\Xi^{\ast}D$} & $\left[\sim3345.0\right]_{V}$ & $\Xi_{c}^{\ast}\bar{K},\Omega_{c}\pi^{+}\pi^{-}\pi^{0}$\tabularnewline
$1\left(\frac{3}{2}^{-}\right)$ &  & $-$ & $-$ &  & $-$ & $-$\tabularnewline
\hline 
$0\left(\frac{1}{2}^{-}\right)$ & \multirow{6}{*}{$\Xi^{\ast}\bar{D}^{\ast}$ $[3542.0]$} & $\left[3541.5_{-2.2}^{+0.4}\right]_{B}^{\dagger}$ & $\Lambda\bar{D}_{s}^{(\ast)},\Xi\bar{D}$ & \multirow{6}{*}{$\Xi^{\ast}D^{\ast}$} & $-$ & $-$\tabularnewline
$0\left(\frac{3}{2}^{-}\right)$ &  & $\left[3541.8_{-0.9}^{+0.1}\right]_{B}^{\dagger}$ & $\Lambda\bar{D}_{s}^{\ast},\Xi\bar{D}\pi$ &  & $-$ & $-$\tabularnewline
$0\left(\frac{5}{2}^{-}\right)$ &  & $\left[3541.9_{-1.4}^{+0.0}\right]_{V}^{\dagger}$ & $\Lambda\bar{D}_{s}^{(\ast)},\Xi\bar{D}$ &  & $\left[3539.3,3539.6\right]_{B}$ & $\Xi_{c}\bar{K},\Xi D,\Omega_{c}\omega$\tabularnewline
$1\left(\frac{1}{2}^{-}\right)$ &  & $-$ & $-$ &  & $\left[3536.7,3541.9\right]_{V}^{\dagger}$ & $\Xi_{c}\bar{K},\Xi D,\Omega_{c}\pi$\tabularnewline
$1\left(\frac{3}{2}^{-}\right)$ &  & $-$ & $-$ &  & $\left[3093.2,3519.9\right]_{V}$ & $\Xi_{c}^{\ast}\bar{K},\Omega_{c}^{\ast}\pi$\tabularnewline
$1\left(\frac{5}{2}^{-}\right)$ &  & $-$ & $-$ &  & $-$ & $-$\tabularnewline
\bottomrule[0.8pt]
\end{tabular}
}
\end{table*}

\subsection{$\Sigma^{(\ast)} K^\ast$, $\Xi^{(\ast)} K^\ast$ and $\Sigma^{(\ast)} \bar{K}^\ast$, $\Xi^{(\ast)} \bar{K}^\ast$ systems}

Whether the existence of hidden-charm pentaquarks implies the existence of hidden-strange pentaquarks is an intriguing question. In the early 2000s, Gao \etal~suggested that the QCD van der Waals force, mediated by multi-gluon exchanges, is strong enough to form a bound state in the $\phi N$ system~\cite{Gao:2000az}. Huang \etal~extended this idea within the chiral SU(3) quark model and demonstrated that the $\sigma$-exchange interaction in $\phi N$ combined with the channel coupling of $\Lambda K^\ast$ can produce a quasi-bound state~\cite{Huang:2005gw}. Liu \etal~using constituent quark models~\cite{Liu:2018nse}, and He \etal~via the effective Lagrangian approach~\cite{He:2018plt}, obtained similar results for the $\phi N$ bound state. Lattice QCD calculations also supported the existence of this bound state~\cite{Beane:2014sda}, although it is important to note that the residual strong interaction of $\phi N$ should be very similar to the $J/\psi N$. Yet, the interaction in the $J/\psi N$ system is too weak to form a bound state due to the Okubo-Zweig-Iizuka (OZI) rule, as shown in a recent lattice QCD calculation~\cite{Skerbis:2018lew}.

Furthermore, Oset \etal~obtained several dynamically generated resonances from vector octet and baryon octet interactions~\cite{Oset:2010tof}, such as the isospin-$\frac{1}{2}$ $\Sigma K^\ast$ and the isospin-$0$ $\Xi K^\ast$. Huang \etal~predicted several isospin-$\frac{1}{2}$ resonances in the $\Sigma^{(\ast)}K^{(\ast)}$ systems~\cite{Huang:2018ehi}, taking into account their couplings to open channels. On the other hand, Yang \etal~argued that there are no hidden-strange pentaquarks in the $\Sigma^{(\ast)}K^{(\ast)}$ systems~\cite{Yang:2022uot}. Lebed proposed a search for hidden-strange pentaquarks in the $\phi p$ invariant mass distributions through the singly Cabibbo-suppressed decay $\Lambda_c^+\to\phi p\pi^0$, as well as open-strange ($s\bar{s}uds$) pentaquarks in the decay $\Xi_c^+\to\phi\Lambda\pi^+$~\cite{Lebed:2015dca}. However, Xie \etal~pointed out the difficulty of identifying hidden-strange pentaquarks in the $\Lambda_c^+\to\phi p\pi^0$ decay due to triangle singularities and small phase space~\cite{Xie:2017mbe}. The Belle Collaboration also investigated the contribution of intermediate hidden-strange pentaquarks in the decay $\Lambda_c^+\to\phi p\pi^0$ but found no evidence with the currently available statistics~\cite{Belle:2017tfw}.

In the following analyses, we will explore the possibility of molecular pentaquarks in the $\Sigma^{(\ast)} K^\ast$, $\Xi^{(\ast)} K^\ast$, $\Sigma^{(\ast)} \bar{K}^\ast$, and $\Xi^{(\ast)} \bar{K}^\ast$ systems, as the correspondences shown in Fig.~\ref{fig:correspondences}. The results are presented in Table~\ref{tab:specHK} and summarized in the last row of Fig.~\ref{fig:spectrum3}. We identified five virtual states in the isospin-$\frac{1}{2}$ $\Sigma^{(\ast)}K^\ast$ system and another five virtual states in the isoscalar $\Xi^{(\ast)}K^\ast$ system. These virtual states can be considered as partners of the $P_{\psi}^N$ and $P_{\psi s}^\Lambda$ states, respectively. Additionally, we found several states, such as the $\frac{1}{2}(\frac{3}{2}^-)$ and $\frac{3}{2}(\frac{1}{2}^-)$ states in $\Sigma\bar{K}^\ast$, the $\frac{1}{2}(\frac{5}{2}^-)$, $\frac{3}{2}(\frac{1}{2}^-)$, and $\frac{3}{2}(\frac{3}{2}^-)$ states in $\Sigma^\ast\bar{K}^\ast$, the $0(\frac{1}{2}^-)$ state in $\Xi\bar{K}^\ast$, and the $0(\frac{5}{2}^-)$, $1(\frac{1}{2}^-)$ states in $\Xi^\ast\bar{K}^\ast$, which are strong candidates of the molecular states.

\begin{table*}[!ht]
\centering
\renewcommand{\arraystretch}{1.5}
\caption{The mass spectra and selected decay channels of the molecular pentaquarks in $\Sigma^{(\ast)}K^{(\ast)}$, $\Xi^{(\ast)}K^{(\ast)}$ and $\Sigma^{(\ast)}\bar{K}^{(\ast)}$, $\Xi^{(\ast)}\bar{K}^{(\ast)}$ systems. The notations are the same as those in Table~\ref{tab:specSD}.\label{tab:specHK}}
\setlength{\tabcolsep}{2.8mm}
{
\begin{tabular}{ccccccc}
\toprule[0.8pt]
$I(J^{P})$ & Systems {[}th.{]} & $E_{B}/E_{V}$ & Decay channels & Systems & $E_{B}/E_{V}$ & Decay channels\tabularnewline
\hline 
$\frac{1}{2}\left(\frac{1}{2}^{-}\right)$ & \multirow{4}{*}{$\Sigma K^{\ast}$ $[2085.2]$} & $\left[2080.6_{-8.8}^{+3.7}\right]_{V}$ & $\phi N,\Lambda K,\Sigma K$ & \multirow{4}{*}{$\Sigma\bar{K}^{\ast}$} & $-$ & $-$\tabularnewline
$\frac{1}{2}\left(\frac{3}{2}^{-}\right)$ &  & $\left[2076.2_{-8.8}^{+5.1}\right]_{V}$ & $\phi N,\Lambda K^{\ast},\Sigma^\ast K$ &  & $\left[2067.6,2076.8\right]_{V}$ & $\Xi^{\ast}\pi,\Xi\pi\pi,\Xi\pi^{+}\pi^{-}\pi^{0}$\tabularnewline
$\frac{3}{2}\left(\frac{1}{2}^{-}\right)$ &  & $-$ & $-$ &  & $\left[2083.2,2085.1\right]_{B}^{\dagger}$ & $\Xi\pi,\Sigma\bar{K}$\tabularnewline
$\frac{3}{2}\left(\frac{3}{2}^{-}\right)$ &  & $-$ & $-$ &  & $-$ & $-$\tabularnewline
\hline 
$\frac{1}{2}\left(\frac{1}{2}^{-}\right)$ & \multirow{6}{*}{$\Sigma^{\ast}K^{\ast}$ $[2276.6]$} & $\left[2274.5_{-7.4}^{+2.1}\right]_{V}$ & $\phi N,\Lambda K,\Sigma K$ & \multirow{6}{*}{$\Sigma^{\ast}\bar{K}^{\ast}$} & $-$ & $-$\tabularnewline
$\frac{1}{2}\left(\frac{3}{2}^{-}\right)$ &  & $\left[2273.2_{-5.0}^{+2.5}\right]_{V}$ & $\phi N,\Lambda K^{\ast},\Sigma^\ast K$ &  & $-$ & $-$\tabularnewline
$\frac{1}{2}\left(\frac{5}{2}^{-}\right)$ &  & $\left[2270.2_{-9.4}^{+4.6}\right]_{V}$ & $\phi N,\Lambda K^{(\ast)},\Sigma^{(\ast)}K$ &  & $\left[2276.4,2276.6\right]_{V}$ & $\Xi^{(\ast)}\pi,\Xi\rho(\omega),\Sigma\bar{K}$\tabularnewline
$\frac{3}{2}\left(\frac{1}{2}^{-}\right)$ &  & $-$ & $-$ &  & $\left[2264.4,2276.6\right]_{B}^{\dagger}$ & $\Xi\pi,\Xi\rho,\Sigma\bar{K}$\tabularnewline
$\frac{3}{2}\left(\frac{3}{2}^{-}\right)$ &  & $-$ & $-$ &  & $\left[1821.5,2261.7\right]_{V}$ & $\Xi^{\ast}\pi,\Xi\rho$\tabularnewline
$\frac{3}{2}\left(\frac{5}{2}^{-}\right)$ &  & $-$ & $-$ &  & $-$ & $-$\tabularnewline
\hline 
$0\left(\frac{1}{2}^{-}\right)$ & \multirow{4}{*}{$\Xi K^{\ast}$ $[2210.3]$} & $\left[2192.8_{-12.2}^{+7.7}\right]_{V}$ & $\phi\Lambda,\Xi K$ & \multirow{4}{*}{$\Xi\bar{K}^{\ast}$} & $\left[2180.8,2193.6\right]_{V}$ & $\Omega\pi^{+}\pi^{-}\pi^{0},\Xi\bar{K}$\tabularnewline
$0\left(\frac{3}{2}^{-}\right)$ &  & $\left[2196.4_{-7.8}^{+5.4}\right]_{V}$ & $\phi\Lambda,\Xi K\pi$ &  & $-$ & $-$\tabularnewline
$1\left(\frac{1}{2}^{-}\right)$ &  & $-$ & $-$ &  & $-$ & $-$\tabularnewline
$1\left(\frac{3}{2}^{-}\right)$ &  & $-$ & $-$ &  & $\left[\sim1533.6\right]_{V}^{\sharp}$ & $\Omega\pi$\tabularnewline
\hline 
$0\left(\frac{1}{2}^{-}\right)$ & \multirow{6}{*}{$\Xi^{\ast}K^{\ast}$ $[2425.4]$} & $\left[2418.7_{-11.6}^{+5.1}\right]_{V}$ & $\phi\Lambda,\Xi K$ & \multirow{6}{*}{$\Xi^{\ast}\bar{K}^{\ast}$} & $-$ & $-$\tabularnewline
$0\left(\frac{3}{2}^{-}\right)$ &  & $\left[2416.5_{-7.6}^{+4.6}\right]_{V}$ & $\phi\Lambda,\Xi K\pi$ &  & $-$ & $-$\tabularnewline
$0\left(\frac{5}{2}^{-}\right)$ &  & $\left[2411.7_{-13.4}^{+7.5}\right]_{V}$ & $\phi\Lambda,\Xi K,\Xi K\pi$ &  & $\left[2423.5,2423.8\right]_{V}$ & $\Omega\pi^{+}\pi^{-}\pi^{0},\Xi\bar{K}$\tabularnewline
$1\left(\frac{1}{2}^{-}\right)$ &  & $-$ & $-$ &  & $\left[2378.2,2423.2\right]_{V}$ & $\Omega\pi\pi,\Xi\bar{K}$\tabularnewline
$1\left(\frac{3}{2}^{-}\right)$ &  & $-$ & $-$ &  & $\left[\sim2310.6\right]_{V}^{\sharp}$ & $\Omega\pi$\tabularnewline
$1\left(\frac{5}{2}^{-}\right)$ &  & $-$ & $-$ &  & $-$ & $-$\tabularnewline
\bottomrule[0.8pt]
\end{tabular}
}
\end{table*}

\section{Summary}\label{sec:sum}

In summary, we extend the concept of heavy quark to cover the strange quark for the near-threshold residual strong interactions. Hence we use the \heavy to denote the hadrons with strange and/or charm quarks. Our study focuses on a systematic examination of the mass spectrum of molecular pentaquarks, composed of a {\it heavy} baryon and a {\it heavy} meson. We consider a range of {\it heavy} baryons, including charmed baryons $\Sigma_c^{(\ast)}$, $\Xi_{c}^{(\prime,\ast)}$, double-charm baryons $\Xi_{cc}^{(\ast)}$, and hyperons $\Sigma^{(\ast)}$ and $\Xi^{(\ast)}$. Meanwhile, the considered {\it heavy} mesons include the charmed mesons $D^{(\ast)}$ and $\bar{D}^{(\ast)}$, as well as the strange vector mesons, namely $K^\ast$ and $\bar{K}^\ast$. By combining these components, we generate various pentaquark systems, which can be classified into two groups based on their interaction types, as outlined in Table~\ref{tab:systems}.

Experimental observations strongly suggest that the formation of the hadronic molecules crucially depends on the correlations between light quark pairs within separate hadrons. In our previous work~\cite{Wang:2023hpp}, we have quantitatively analyzed these interactions and demonstrated that the strange quarks inside the hadrons can be approximated as {\it heavy} quarks due to their limited involvement in near-threshold interactions. Consequently, we establish the underlying connections between different systems under the framework of {\it heavy} quark symmetry, as depicted in Fig.~\ref{fig:correspondences}.

Our analysis reveals that the very near-threshold behavior of $P_{\psi s}^\Lambda(4338)$ shares a common cause with the similarly near-threshold $T_{cc}(3875)$. Furthermore, we find that the masses of the molecular candidates in the $\Xi_c\bar{D}^\ast$ system should be closer to the threshold than the currently available experimental data, specifically the $P_{\psi s}^\Lambda(4459)$. 
Through our investigations, we have identified numerous hadronic molecules within the twenty systems listed in Table~\ref{tab:systems}. These molecular states can be considered as counterparts to the $P_{\psi}^N$, $T_{cc}$, $X(3872)$, and $Z_c(3900)$ states, expanding the family of hadronic molecules. 

The study of pentaquark systems with different quark configurations, particularly in the realm of hadronic molecules, has provided valuable insights into the nature of exotic hadrons and has helped to deepen our understanding of the experimental observations in this field. The experimental search for the predicted states in various decay channels continues to be of great importance in advancing our understanding of the underlying physics.

\section*{Acknowledgement}
This work is supported by the National Natural Science Foundation of China under Grants Nos. 12105072, 12305090, 11975033 and 12070131001. B. Wang was also supported by the Youth Funds of Hebei Province (No. A2021201027) and the Start-up Funds for Young Talents of Hebei University (No. 521100221021). This project is also funded by the Deutsche Forschungsgemeinschaft (DFG, German Research Foundation, Project ID 196253076-TRR 110).

\bibliography{refs}
\end{document}